\documentclass[twocolumn]{article}

\usepackage[pdftex]{graphicx}
\graphicspath{{./}} 

\usepackage[T1]{fontenc}
\usepackage{hyperref}
\usepackage{cite}

\usepackage{textcomp}   
\usepackage{booktabs}   
\usepackage{amsmath}
\usepackage{amsthm}
\usepackage{mathtools}
\usepackage{subcaption}
\usepackage[ruled,commentsnumbered]{algorithm2e}
\usepackage{xcolor}         
\usepackage{overpic}
\usepackage{tabularx}
\usepackage{appendix}
\usepackage{ellipsis}

\makeatletter
\renewcommand*{\mathellipsis}{%
	\mathinner{%
		\kern\ellipsisbeforegap%
		{\ldotp}\kern\ellipsisgap%
		{\ldotp}\kern\ellipsisgap%
		{\ldotp}\kern\ellipsisaftergap%
	}%
}
\renewcommand*{\dotsb@}{%
	\mathinner{%
		\kern\ellipsisbeforegap%
		{\cdotp}\kern\ellipsisgap%
		{\cdotp}\kern\ellipsisgap%
		{\cdotp}\kern\ellipsisaftergap%
	}%
}
\renewcommand*{\@cdots}{%
	\mathinner{%
		\kern\ellipsisbeforegap%
		{\cdotp}\kern\ellipsisgap%
		{\cdotp}\kern\ellipsisgap%
		{\cdotp}\kern\ellipsisaftergap%
	}%
}
\renewcommand*{\ellipsis@default}{%
	\ellipsis@before
	\kern\ellipsisbeforegap
	.\kern\ellipsisgap
	.\kern\ellipsisgap
	.\kern\ellipsisgap
	\ellipsis@after\relax}
\renewcommand*{\ellipsis@centered}{%
	\ellipsis@before
	\kern\ellipsisbeforegap
	.\kern\ellipsisgap
	.\kern\ellipsisgap
	.\kern\ellipsisaftergap
	\ellipsis@after\relax}
\AtBeginDocument{%
	\DeclareRobustCommand*{\dots}{%
		\ifmmode\@xp\mdots@\else\@xp\textellipsis\fi}}
\def\ellipsisgap{.05em}
\def\ellipsisbeforegap{0em}
\def\ellipsisaftergap{0em}
\makeatother

\DeclareMathOperator*{\argmin}{arg\,min}

\hypersetup{
	colorlinks,
	linkcolor={red!50!black},
	citecolor={blue!50!black},
	urlcolor={blue!80!black}
}

\begin{document}

\title{Optimally Fast Soft Shadows on Curved Terrain with Dynamic Programming and Maximum Mipmaps}

\author{Dawoon Jung$^{1}$, Fridger Schrempp$^{2}$
        and Seunghee Son$^{1}$
    \\
    {\parbox{\textwidth}{\centering \vspace{1em}$^1$Korea Aerospace Research Institute (KARI), Daejeon, Korea\\
            \{dwjung|seunghee\}@kari.re.kr
            \\
            $^2$Deutsches Elektronen-Synchrotron (DESY), Hamburg, Germany\\
            fridger.schrempp@desy.de
        }
    }
}


\maketitle
\begin{abstract}
    We present a simple, novel method of efficiently rendering ray cast soft shadows on curved terrain by using dynamic programming and maximum mipmaps to rapidly find a global minimum shadow cost in constant runtime complexity. Additionally, we apply a new method of reducing view ray computation times that pre-displaces the terrain mesh to bootstrap ray starting positions. Combining these two methods, our ray casting engine runs in real-time with more than 200\% speed up over uniform ray stepping with comparable image quality and without hardware ray tracing acceleration. To add support for accurate planetary ephemerides and interactive features, we integrated the engine into celestia.Sci, a general space simulation software. We demonstrate the ability of our engine to accurately handle a large range of distance scales by using it to generate videos of lunar landing trajectories. The numerical error when compared with real lunar mission imagery is small, demonstrating the accuracy and efficiency of our approach.
\end{abstract}  
\section{Introduction}
Modeling soft shadows accurately and efficiently is important for interactive rendering of long shadows cast by terrain on curved, planetary surfaces. In our case, we are developing an interactive simulator that must be able to generate reference imagery for testing autonomous landing at the lunar poles. Potential landing sites in these regions are perpetually in twilight, and low sun elevation angles lead to soft penumbrae that are nearly $1/7$th the total length of shadows that are themselves tens of kilometers long.

Soft shadows can be computed in a single render pass using ray casting. While several acceleration techniques for computing ray-terrain intersections exist\cite{Oliveira2005,Tevs2008}, such techniques typically are problematic when ray casting the surface of a curved base surface such as a planet, require expensive precomputation, or result in inaccuracies. Moreover, tracing rays for soft shadows cannot be terminated early (\autoref{subsec:shadows}), requiring a large number of iterations of slow texture lookups that dominate the render cost.

In this work, we describe how to overcome the slow speed of shadow ray casting and also introduce a simple optimization for view rays to efficiently and accurately render planet-scale, curved, rugged, and heavily-shadowed terrain with soft shadows in a real-time, interactive application running on previous-generation hardware. We demonstrate and validate our results using mainly scientific data of the Moon, but later we briefly show how our technique generalizes to bodies with atmospheres such as Mars and Pluto. 

In essence, we recognized that the problem of ray casting soft shadows is equivalent to finding the global minimum of a cost function in as few iterations as possible where the cost is proportional to the shadow ray height above the terrain surface. We found that \emph{dynamic programming}--a recursive optimization technique--works well for this case. Our technique relies on a maximum mipmap \cite{Tevs2008} containing maximum height values in successive mip levels that is generated in real-time only for the visible terrain area and when the view frustum changes sufficiently, in order to guarantee that height above the terrain is minimized for a given subproblem.

We also show that view ray iteration step count can be reduced by a factor of nearly 100 for an equivalent or better intersection accuracy by pre-displacing the terrain mesh in the vertex shader.

To summarize our main contributions again:
\begin{enumerate}
    \item \textbf{Fast, accurate}, real-time method of computing soft shadows on spherical geometry using dynamic programming and maximum mipmaps
    \item Efficient \textbf{hybrid} vertex displacement and view ray casting
    \item \textbf{Scalable} rendering architecture that can handle planetary distance scales
\end{enumerate}

\begin{figure*}[t]
    \centering
    \begin{overpic}[width=0.7\textwidth]{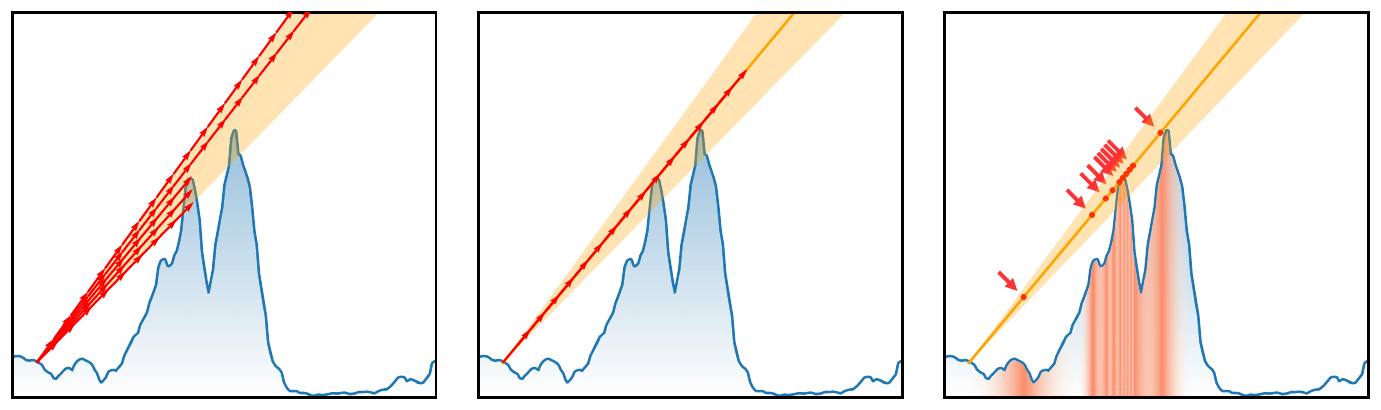}
        \put(22.7,22.5){\color{orange}{\small\textsf{area light}}}
        \put(3,25){(a)}\phantomsubcaption{\label{subfig:raydistrib}}
        \put(37,25){(b)}\phantomsubcaption{\label{subfig:rayuniform}}
        \put(71,25){(c)}\phantomsubcaption{\label{subfig:rayfast}}
    \end{overpic}\caption{Comparison of shadow ray tracing methods. (\subref*{subfig:raydistrib}) Distributed ray tracing requires tracing multiple, expensive rays. (\subref*{subfig:rayuniform}) Uniform stepping with analytic occluded light area calculation requires only marching along direct line of sight but many samples are still wasted. (\subref*{subfig:rayfast}) Our method. Horizontal shading represents maximum mipmap samples, filtered to prevent aliasing. Arrows indicate points along the shadow ray where samples are required. Early samples occur at high mip levels, allowing the algorithm to skip large swaths of the total interval.}
    \label{fig:shadowprofile}
\end{figure*}

\subsection{Overview}

\autoref{sec:related} discusses research related to this work.
\autoref{sec:raycasting} details our soft shadow algorithm.
\autoref{sec:viewraycasting} describes our hybrid view ray casting method.
\autoref{sec:celestia} discusses how our ray casting engine fits into celestia.Sci.
\autoref{sec:results} presents our results, and
\autoref{sec:conclusion} gives concluding remarks.

\section{Related Work}
\label{sec:related}
Terrain is often rendered using displacement mapping, a technique that usually involves displacing the vertices of a polygonal mesh or ray casted surface based on elevation values looked up from a height field texture. Szirmay-Kalos and Umenhoffer \cite{Szirmay-Kalos2008} provide a comprehensive review of displacement mapping methods.

Self-shadowing on terrain refers to shadows cast on terrain by the terrain itself, and is typically computed using shadow maps in the case of polygonal displacement mapping, or ray casting (hybrid methods also exist, e.g., grid tracing \cite{Musgrave1990} which casts rays against a polygonal grid). Ray marching is a variant of ray casting where the ray is marched in discrete steps.

When computing shadows using ray casting, a \emph{view ray} from the viewer to the terrain surface is cast, and then a \emph{shadow ray} or bundle of rays is shot from the terrain intersection point to each light source to determine the shadow intensity. Soft shadows are shadows whose edges are softened due to area lights having non-zero area, e.g., in the case of the Sun. Completely shadowed portions are defined as umbra, and soft areas are defined as penumbra. The reference method of tracing soft shadows is distributed ray tracing \cite{Cook1984} (\autoref{subfig:raydistrib}), but this requires casting many rays to each light source with predictably slow performance.

Recent techniques aim to reduce ray casting cost by requiring less ray samples. These include horizon mapping techniques that compute the horizon silhouette \cite{Snyder2008,Nowrouzezahrai2009,Timonen2010} and cone step mapping\cite{Chang2009} in a cone step map normally used for accelerating ray-surface intersections is repurposed to compute average visibility in a limited set of directions. However, these methods are limited by slow precomputation times and coarse sampling.

Uniform sampling\cite{Tatarchuk2006,Drobot2018} takes constant-size steps along a single ray until the ray exits the terrain volume. No precomputation is required, but high-frequency terrain features might be missed unless sampled at sufficiently high rates (\autoref{subfig:rayuniform}) and even then the computational complexity is $\mathcal{O}(N)$ where time to trace the entire shadow interval increases linearly with number of steps $N$.

Why then are existing ray-cast soft shadowing methods so unsatisfactory? Certainly many optimized ray-surface intersection methods exist; the reader is referred to maximum mipmaps, relaxed cone step mapping\cite{Policarpo2007} (which must not be confused with ref \cite{Chang2009} where cone sectors encode horizon information), and precomputed safety shapes\cite{Baboud2012} for the state-of-the-art. However soft shadows cannot be computed using these methods on their own, because they only solve for the distance to an intersection.

However, soft shadow computation is a fundamentally different problem than ray-surface intersection. The goal of soft shadow computation is \emph{not} to find the distance to an intersection. Instead, it is to find the \textbf{occluded fraction of an area light}. Indeed, within the solid angle subtended by the area light there may be many shadow rays which do not intersect a surface.

In this work, we show that calculating a soft shadow can be cast as a \emph{global minimization} problem. Global minimization is a well-studied area\cite{Burke2014}. Classical methods such as dynamic programming\cite{Bertsekas2005} and branch and bound are useful when the problem can be divided into sub-problems\cite{Dowsland2014}, while heuristic methods such as simulated annealing and tabu search do not strongly guarantee optimality but can be efficient for large problems.

\autoref{subfig:rayfast} illustrates our method. We start with the same strategy as uniform sampling by tracing a single ray trajectory toward the center of an area light source, but use dynamic programming guided by a maximum mipmap generated cheaply in real-time to find a global minimum shadow cost along the trajectory in $\mathcal{O}(\log N)$ runtime in the worst case, or $\mathcal{O}(1)$ in practice. The maximum mipmap generation and optimal ray casting run on the GPU in standard OpenGL vertex and fragment shaders
at more than 3 times the frame rate of uniform ray stepping with comparable image quality and without hardware ray tracing acceleration.

\section{Our Method}
\label{sec:raycasting}

This section describes the method we use to compute soft shadows accurately and efficiently.

\subsection{Soft Shadow Theory}
\label{subsec:shadows}

\begin{figure}[ht]
	\centering
	\includegraphics[width=0.48\textwidth]{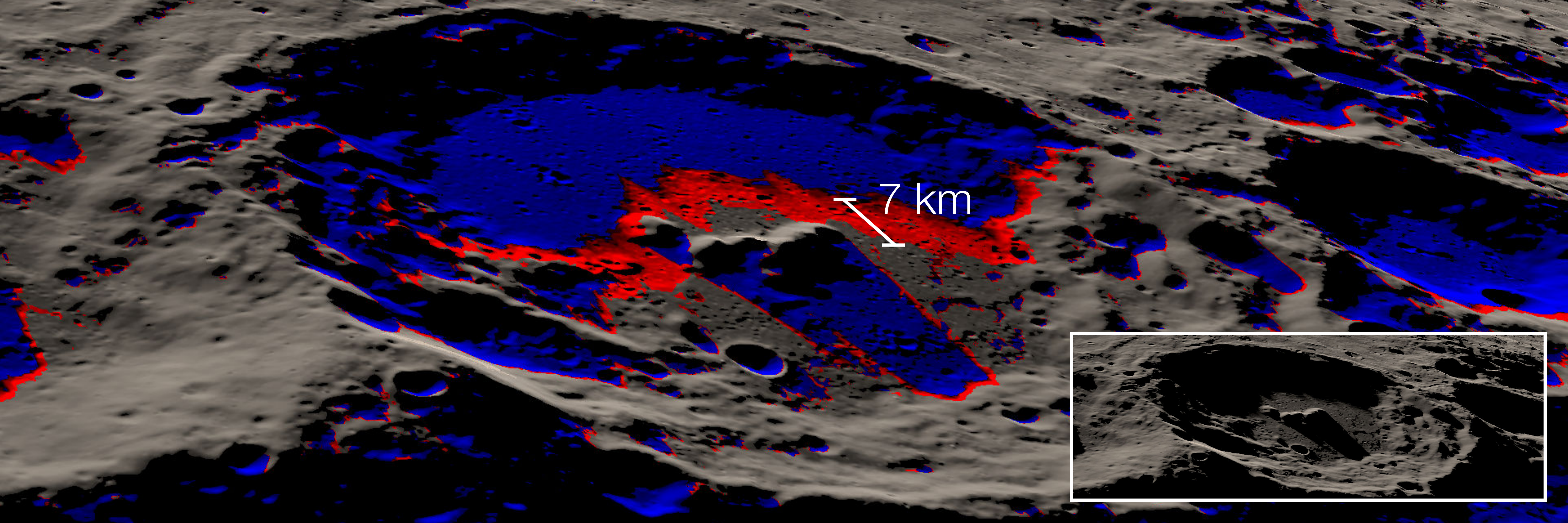}
	\caption{%
		Color coded illustration of long penumbrae cast by lunar crater rims in our renderer (inset: final render). Sunlight is shining from the top left. 
		Umbrae (blue) can reach over 30 km in length, and penumbrae (red) over 5 km or nearly 1/7th the total shadow length in this scene. By comparison, maximum height field resolution is 30 m/texel.}
	\label{fig:penumbrademo}
\end{figure}

Modeling soft shadows is important to us, because low solar elevation angles near the lunar poles create shadows that are tens of kilometers in length and penumbrae that are a significant fraction of the length (\autoref{fig:penumbrademo}).

\begin{table}[htb]
	\caption{Table of symbols.}%
	\label{tab:symbols}
	\centering
	\begin{tabularx}{\linewidth}{|c|X|}
		\hline
		$\mathbf{p}$ & View ray intersection point \\
		$\mathbf{\hat{L}}$ & Shadow (light) ray unit vector \\
		$\mathbf{R}_\text{obj}$ & Object space position vector of tip of ray \\
$\mathbf{R}_\text{tex}$ & Texture space position vector of tip of ray \\
		$h$ & Terrain height in $[0, 1]$ increasing from bottom to top \\
		$\max h$ & Maximum $h$ within a maximum mipmap texel \\
		$H$ & Height of tip of ray \\
		$\Delta\max h_k$ & Difference between $\max h$ and $H$ \\
		$T$ & Length of ray covering one maximum mipmap texel \\
		$t$ & Distance traveled by shadow ray normalized to $[0, T]\rightarrow[0, 1]$ \\
		$N$ & Height field size \\
		$N'$ & Total number of mipmap levels \\
		$m$ & Mip level \\
		$\Delta M$ & Size of mipmap texel \\
		$k$ & Current iteration step \\
		$J\text{*}$ & Optimal cost over $\{0, 1,\ldots,k\}$ \\
		\hline
	\end{tabularx}
\end{table}



\begin{figure}[htbp]
    \centering
    \begin{subfigure}{0.23\textwidth}
        \includegraphics[width=1.0\textwidth]{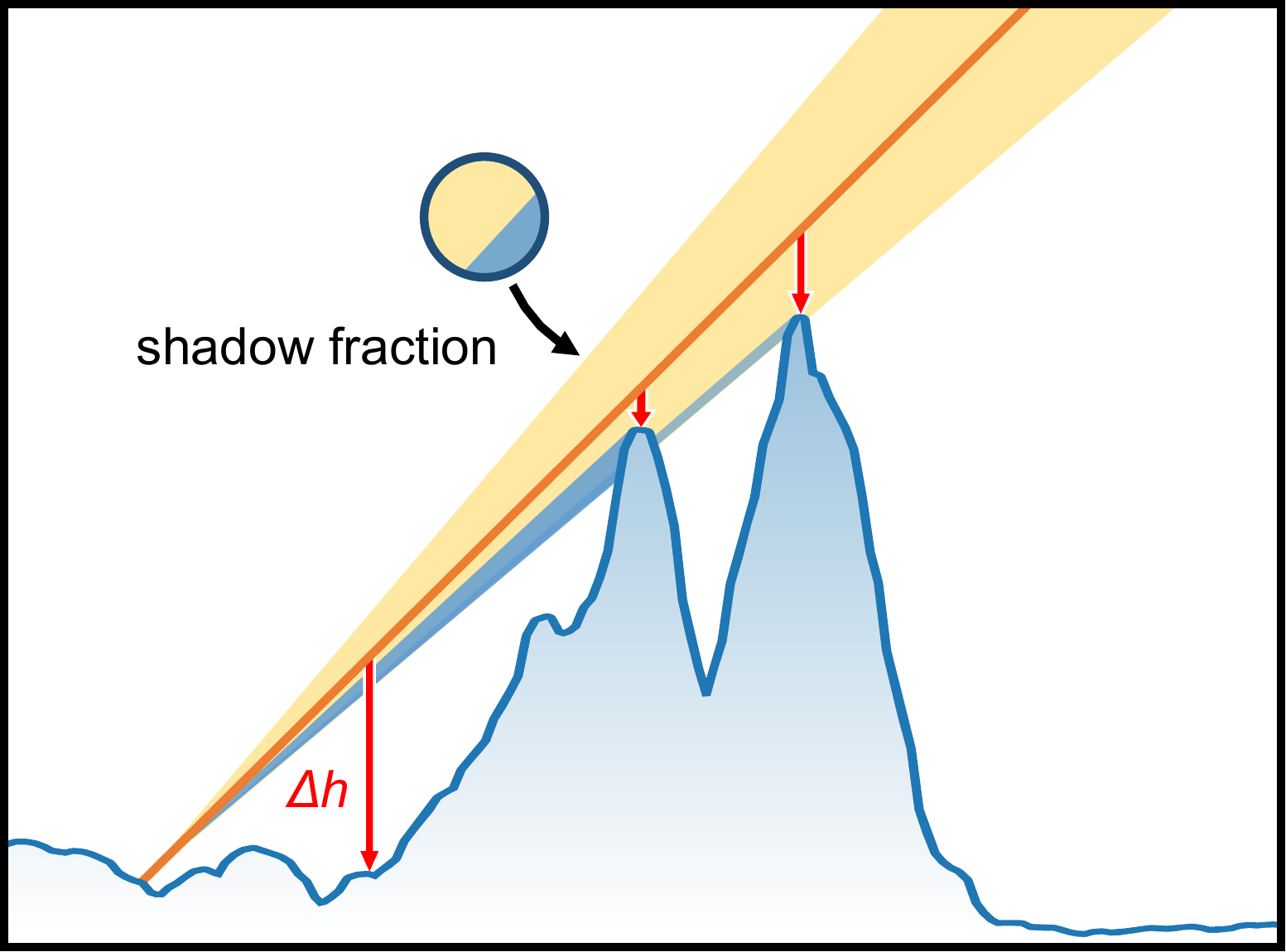}\caption{Penumbra geometry}\label{subfig:penumbrageom}
    \end{subfigure}
    \begin{subfigure}{0.23\textwidth}
        \includegraphics[width=1.0\textwidth]{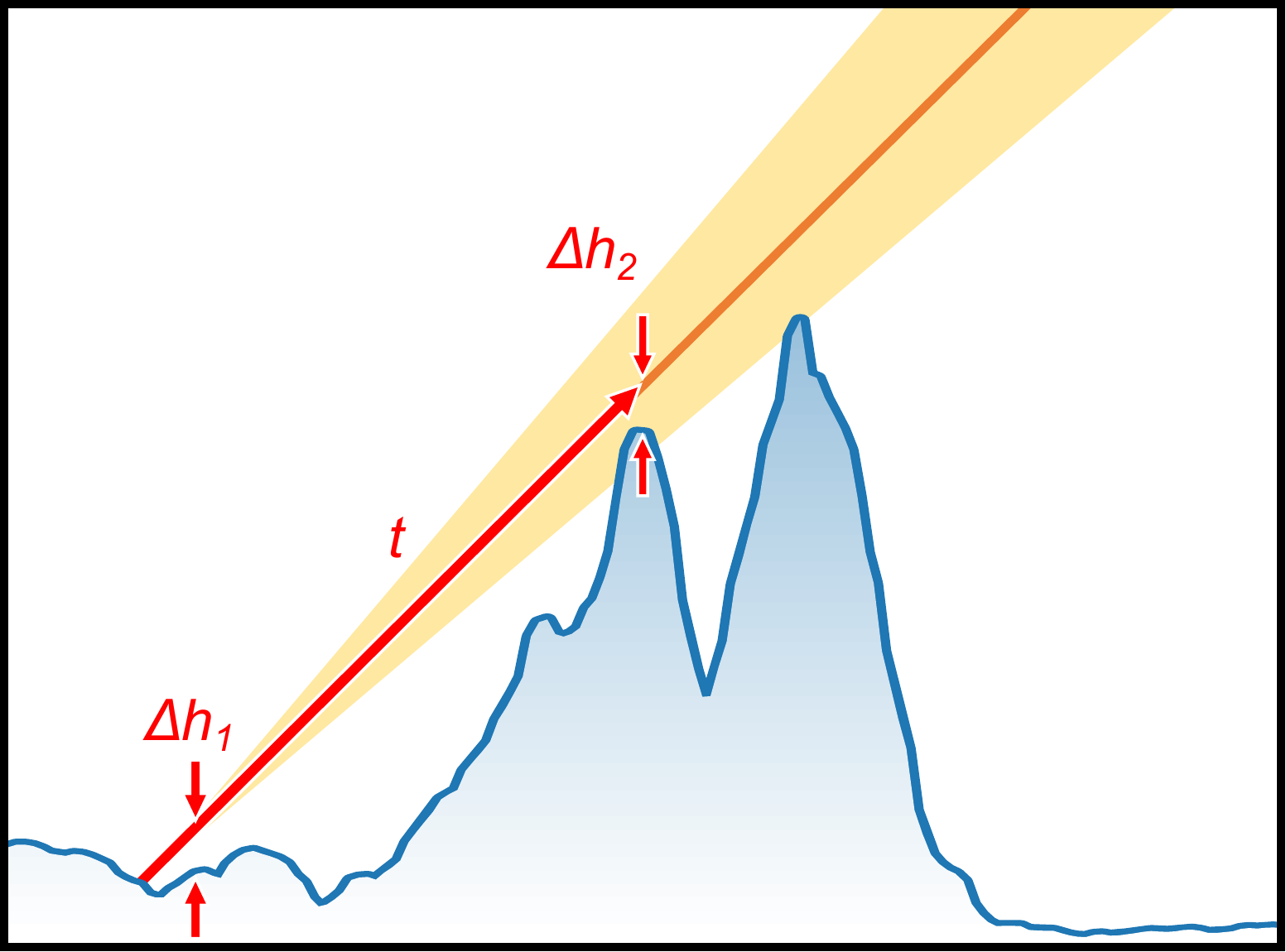}\caption{Penumbra cost}\label{subfig:penumbracost}
    \end{subfigure}\caption{Penumbra geometry and cost}
    \label{fig:penumbra}
\end{figure}

Our method computes the fraction $s$ of an area light that is shadowed by terrain. It does this by shooting a single shadow ray $\mathbf{R}_{\text{obj}}=\mathbf{p}+\mathbf{\hat{L}}t$ from a view ray intersection point $\mathbf{p}$ toward the geometric center of the area light, and computing the height difference between the shadow ray and terrain $\Delta h_k$ at strategic points. We observe in \autoref{subfig:penumbrageom} that the shadow fraction $s_k$ is larger when $\Delta h_k$ is small. However, small $\Delta h_k$ also always occur trivially just after the view ray intersection point ($\Delta h_1$ in \autoref{subfig:penumbracost}). Instead, the shadow fraction should also take into account the distance travelled by the shadow ray $t_k$ in the following manner \cite{Tatarchuk2006,Tamas2016}:
\begin{equation}\label{eq:shadowfrac}
s_k = S\left(\frac{\Delta h_k}{t_k}\right)
\end{equation}

where $S$ is a function relating $\Delta h/t$ to the sun disc occlusion fraction as will be described in \autoref{subsec:sunocclusion}.
 $\Delta h / t$ can be interpreted in terms of the unobscured solid angle of a cone of light with apex at the start of the shadow ray.
 
As minimizing $\Delta h / t$ will minimize $s$, the optimal cost $J\text{*}(s\text{*})$ might appear to be:
\begin{equation}
J\text{*}(s\text{*})=\min\frac{\Delta h}{t}
\end{equation}
where $s\text{*}=\max(s)$ (for brevity we will omit the $(s\text{*})$ from henceforth). However, the cost function that we should use is:
\begin{equation}\label{eq:cost}
J\text{*} = \min\Delta h
\end{equation}
Only the numerator should be minimized, otherwise $\Delta h / t$ can be trivially minimized by arbitrarily increasing $t\rightarrow\infty$.


Since there may be any number of `peaks' along the shadow ray with small but non-negative $J_k=\Delta h_k$, the entire length of the shadow ray from $p$ to the point where the ray exits the terrain volume must be considered. Next we show how our method solves this global optimization problem using dynamic programming in a single-pass GPU fragment shader.

\subsection{Dynamic Programming}
\label{subsec:dynamicprog}

As discussed in \autoref{sec:related}, we choose dynamic programming to solve the shadow ray global minimization problem. Dynamic programming is an algorithm for solving global minimization problems recursively in cases where the cost to optimize a step depends on previous steps \cite{Bellman1954}. The solution will be optimal if each step is optimal. In other words, if $J\raisebox{0.1em}{\text{*}}_{\hspace{-.5em}k}$
is optimal, then $J\raisebox{0.1em}{\text{*}}_{\hspace{-.5em}k+1}=\min J\raisebox{0.1em}{\text{*}}_{\hspace{-.5em}k}$ will be optimal. The task then, is to find a way to generate an efficient \emph{policy} set $\pi=\{J\raisebox{0.1em}{\text{*}}_{\hspace{-.5em}0}, J\raisebox{0.1em}{\text{*}}_{\hspace{-.5em}1}, \ldots, J\raisebox{0.1em}{\text{*}}_{\hspace{-.5em}k}\}$ that will satisfy optimality while guaranteeing that the policy covers the entire shadow ray domain. We show that maximum mipmaps can be used to achieve this.

\noindent\textbf{Proof of optimality}
	The ray parameter $t\text{*}$ resulting in the optimal shadow cost can be trivially defined as:
	\begin{equation}
		t\text{*} = \argmin\limits_{t}\left\{\min\Delta h(t)\right\} 
	\end{equation}
	We will now show how to derive an efficient variant of this that uses dynamic programming to run in $\mathcal{O}(\log N)$ time.

	$t\text{*}$ can be calculated by the following dynamic programming method:
	\begin{align*}
		t\raisebox{0.1em}{\text{*}}_{\hspace{-.5em}k} =
		\begin{cases}
			 \qquad\qquad 1 & k=0\\
			\argmin\limits_{t}\left\{\min\Delta\max h(t\raisebox{0.1em}{\text{*}}_{\hspace{-.5em}k+1}+2^{-k})\right\}&k>0
		\end{cases}
	\end{align*}
	where each step $k$ represents a subdivision by half of step.

	If $k=0$ (the base case), $t=1$ because the ray covers the entire texture. For subsequent $k>0$, we first write $t\raisebox{0.1em}{\text{*}}_{\hspace{-.5em}k}=\argmin\{\min\Delta h(t_{k,i})\}=\argmin\{\min\{\Delta h(t_{k,0}), \Delta h(t_{k,1}), \dots\}\}$ where $i=\{0,1,\dots\}$ and $t_{k,i}$ represents values of $t$ over the interval covering all possible texels intersected by the ray.

	$\Delta h(t_{k,i})$ can be expressed in terms of subsequent subdivisions at $k+1$:
	\begin{align}
		\min\Delta h(t_{k,i})&=\min\{H(t_{k,i}) - h(t_{k,i})\} \nonumber\\
		&=\min\{H(t_{k,i}) - \max\{h(t_{k+1,2i}), \nonumber\\
		&\qquad\qquad h(t_{k+1,2i+1}),\dots\}\} \nonumber\\
		&=\min\Delta\max h(t_{k+1}) \label{eq:optimumk2}
	\end{align}
	where we write the height of the ray tip in texture space as a function of $t_{k,i}$ as $H(t_{k,i})\equiv\mathbf{R}_{\text{tex},k,i}^h$ where $\mathbf{R}_{\text{tex},k,i}=\frac{1}{2}\mathbf{R}_{\text{obj},k,i}^{xy}/\mathbf{R}_{\text{obj},k,i}^{z}+\frac{1}{2}$ for spherical terrain. To see why $h(t_{k,i})$ must take the maximum of $\{h(t_{k+1,2i}), h(t_{k+1,2i+1}),\dots\}$, first note that the ray in the interval covered by the $i$th texel at mip level $k$ necessarily intersects texels $2i, 2i+1, \dots$ at mip level $k+1$. If $\Delta h(t_{k,i})$ were to minimize $t_k$, then none of $\Delta h(t_{k+1})$ in the interval covered by the ray in the $i$th texel can be greater than $\Delta h(t_{k,i})$ since otherwise the interval would not minimize $t_k$. Additionally, $H(t_{k,i})$ is monotonically increasing with respect to the terrain (otherwise it intersects the terrain and is trivially shadowed) so $h(t_{k,i})$ must be the largest possible in order to minimize $\Delta h(t_{k,i})$.

	Additionally, when going to the next subdivision level, texel centers become offset by $2^{-k}$. So $t\raisebox{0.1em}{\text{*}}_{\hspace{-.5em}k} = \argmin\left\{\min\Delta\max h(t\raisebox{0.1em}{\text{*}}_{\hspace{-.5em}k+1}+2^{-k})\right\}$ as claimed.

\noindent\textbf{Runtime complexity}
	To show that $t\text{*}$ can be calculated in  $\mathcal{O}(\log N)$ time,
	note that each subdivision equates with a mip level and the total number of mip levels is $\log N$ where $N$ is the maximum dimension (width or height in pixels) of the height field. The total number of steps required at each mip level is constant (up to three in our case). Thus the total runtime complexity is bounded by $\log N$.

Empirically however, we find that it is not necessary to trace all $\log N$ mip levels but only a constant $N'<\log N$ every time, if we always use a texture size that is on the order of the display size (e.g.,  $N'=5$ for $N=[1024, 2048]$) and noting that most shadows do not cover the entire length of the texture. In this case, runtime complexity is \emph{constant}, i.e., $\mathcal{O}(1)$.

\noindent\textbf{Maximum mipmaps}
	On a 1-dimensional terrain, each $\max h(t\raisebox{0.1em}{\text{*}}_{\hspace{-.5em}k+1}+2^{-k})$ can be obtained by sampling from a maximum mipmap which stores the maximum values of the corresponding collection of height field texels in successively decreasing mip levels $m_k=\{N'-1, \ldots, 1, 0\}$.

	To prove this, first we define 1-dimensional terrain as terrain that has minimal variation in the direction perpendicular to the shadow ray. In practice this is not always true but we will later show that the artifacts resulting from this approximation can be effectively mitigated as described in the section on long shadows in \autoref{subsec:scalablerender}.

	By \autoref{eq:optimumk2}, each $h(t_{k,i})=\max{h(t_{k+1,2i}),\dots}$ where $h(t_{k+1,2i}),\dots$ are the heights of texels at mip level $k+1$ in the interval entirely covered by the texel at mip level $k$. But this is just the definition of a maximum mipmap texture.

\noindent\textbf{Texel traversal}
We initially set a shadow ray of length $T$ to correspond to $\Delta\mathbf{M}$, the vector projected by the ray into texture space of a maximum mipmap texel at mip level $m_0=N'-1$, aiming to cover one texel (see \nameref{def:deltaT} in the Appendix for a complete definition).

\begin{figure}[htbp]
	\centering
	\begin{subfigure}{0.13\textwidth}
		\includegraphics[width=\textwidth]{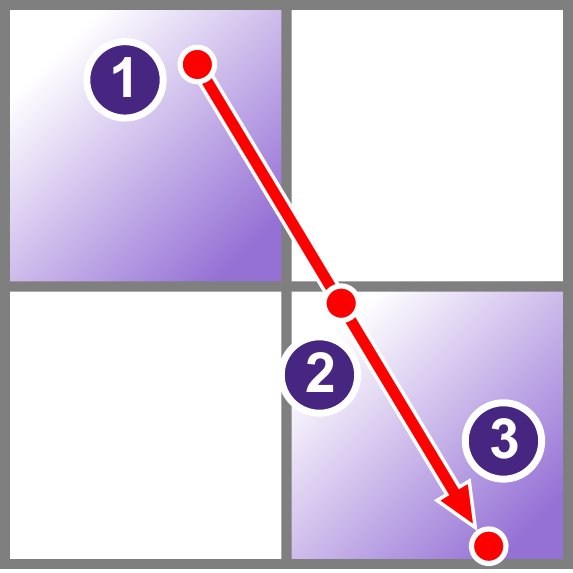}\caption{\label{subfig:ddacanon}}
	\end{subfigure}
	\begin{subfigure}{0.13\textwidth}
		\includegraphics[width=\textwidth]{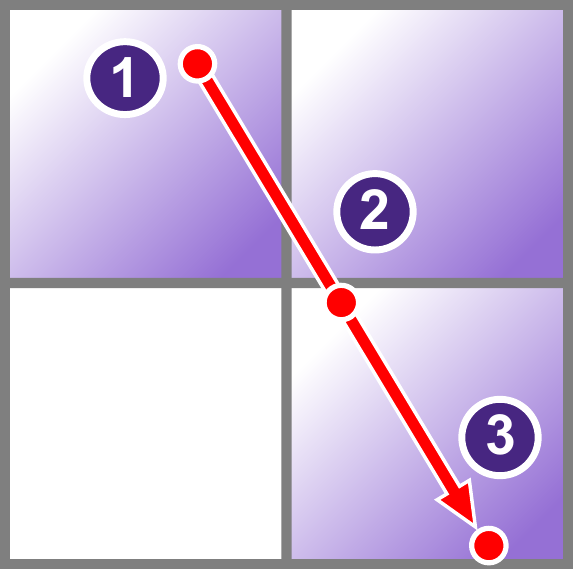}\caption{\label{subfig:dda}}
	\end{subfigure}
	\caption{Texels intersected by ray of length $T$. (\subref*{subfig:ddacanon}) Canonical DDA sometimes misses a texel. (\subref*{subfig:dda}) Up to three texels of size $\Delta M/2$ could be intersected; we visit them all and find the texel that minimizes $J_k$.%
	}
	\label{fig:streakdda}
\end{figure}

\begin{figure}
	\centering
  	\setlength{\unitlength}{0.16\textwidth}
	\begin{subfigure}{\unitlength}
		\begin{picture}(1,1)
		\put(0,0){\includegraphics[trim=17cm 9cm 14cm 3cm, clip=true,height=\textwidth]{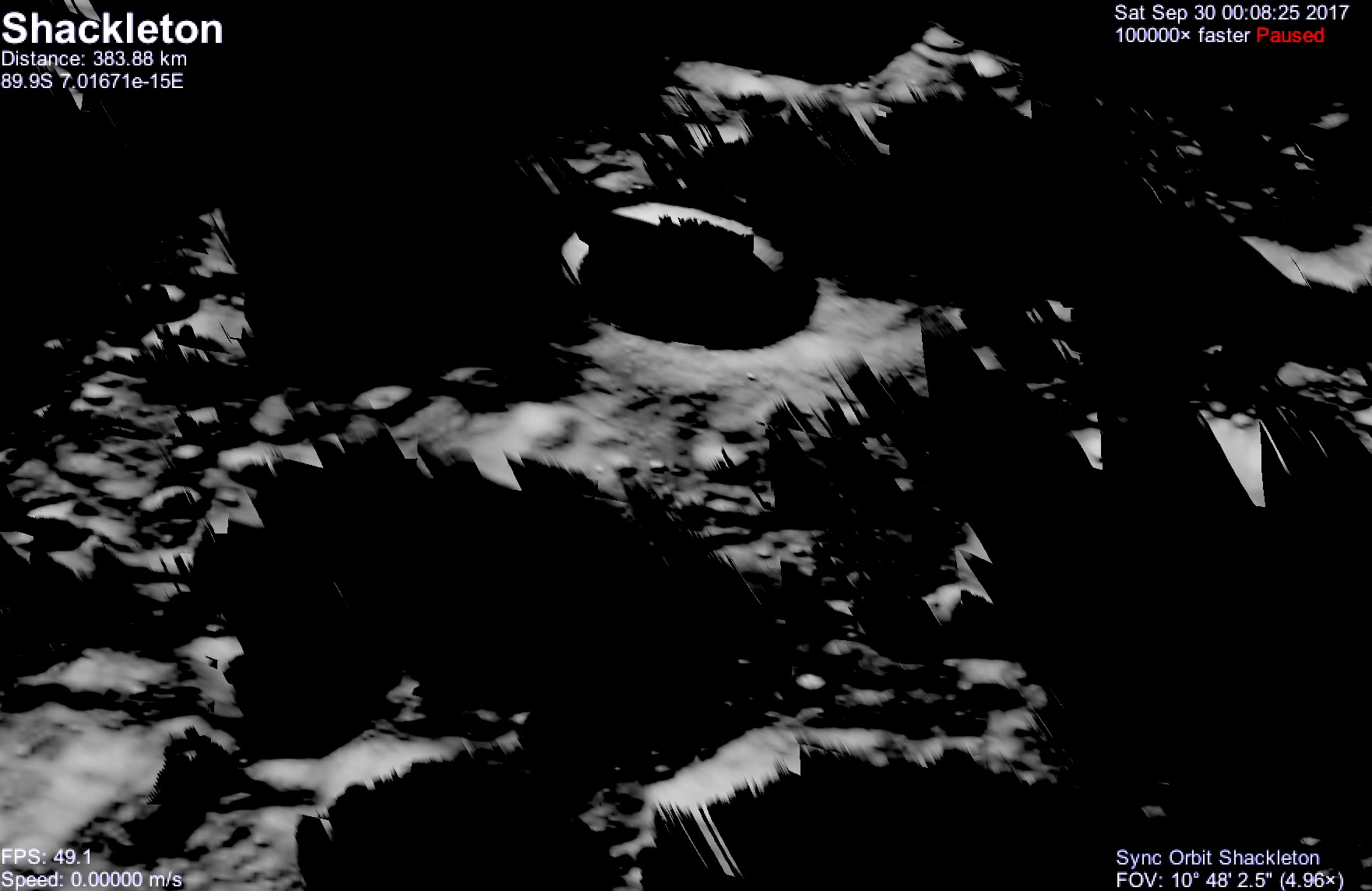}}
		\put(0.05,.8){\color{white}{(a)}}
		\end{picture}\phantomcaption{\label{subfig:streaks-shackleton}}
	\end{subfigure}
	\hspace{0.05\unitlength}
	\begin{subfigure}{\unitlength}
		\begin{picture}(1,1)
		\put(0,0){\includegraphics[trim=17cm 9cm 14cm 3cm, clip=true,height=\textwidth]{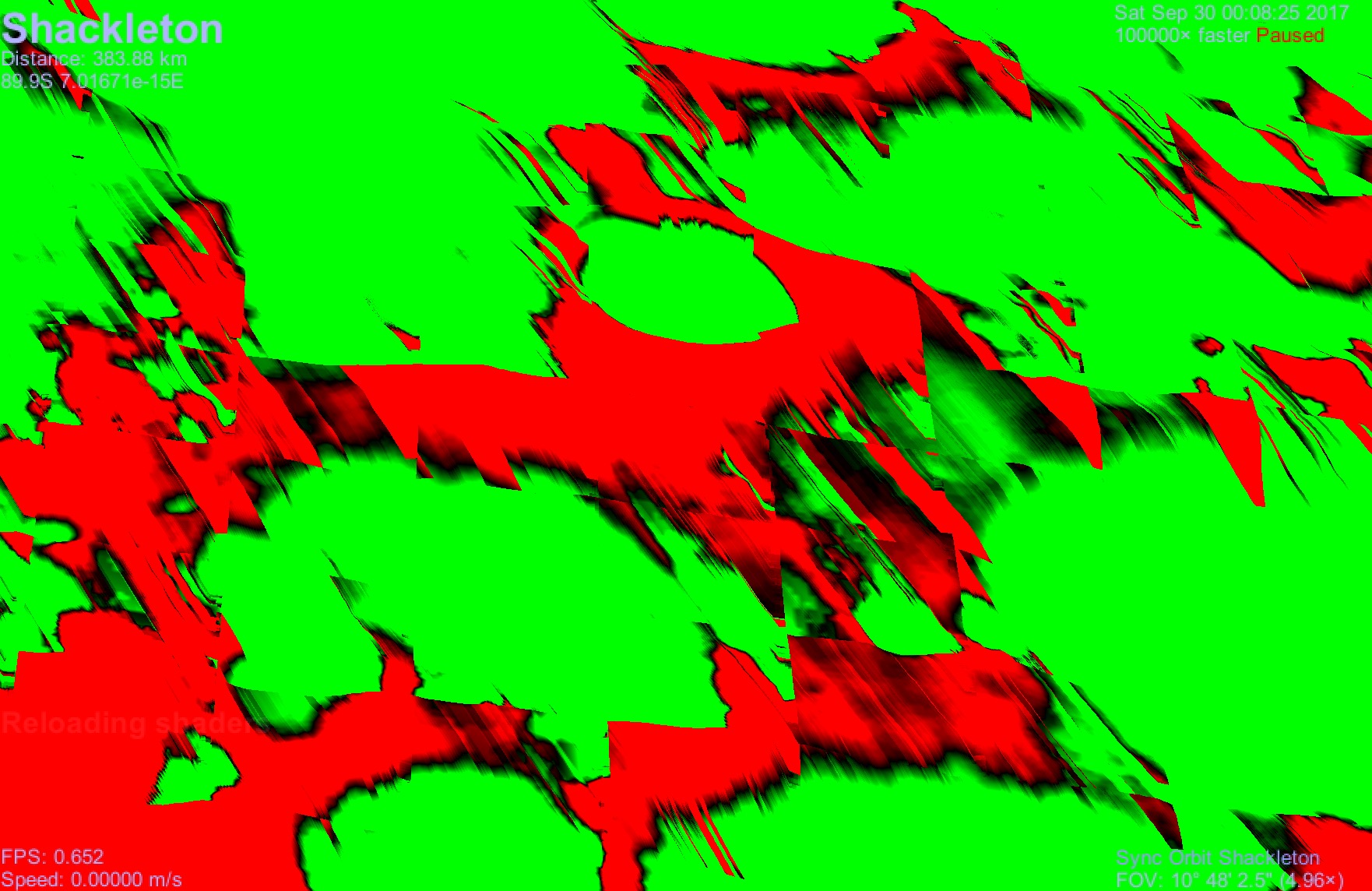}}
		\put(.05,.8){(b)}
		\end{picture}\phantomcaption{\label{subfig:streaks-J}}
	\end{subfigure}
	\hspace{0.07\unitlength}
  	\begin{subfigure}{0.3333\unitlength}
		\begin{picture}(.3333,1)
		\put(0,0){\includegraphics[height=\unitlength]{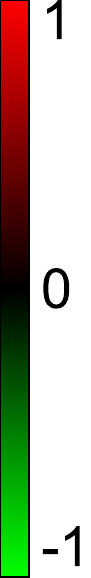}}
		\end{picture}
	\end{subfigure}\\
	\begin{subfigure}{\unitlength}
		\begin{picture}(1,1)
		\put(0,0){\includegraphics[trim=17cm 9cm 14cm 3cm, clip=true,height=\textwidth]{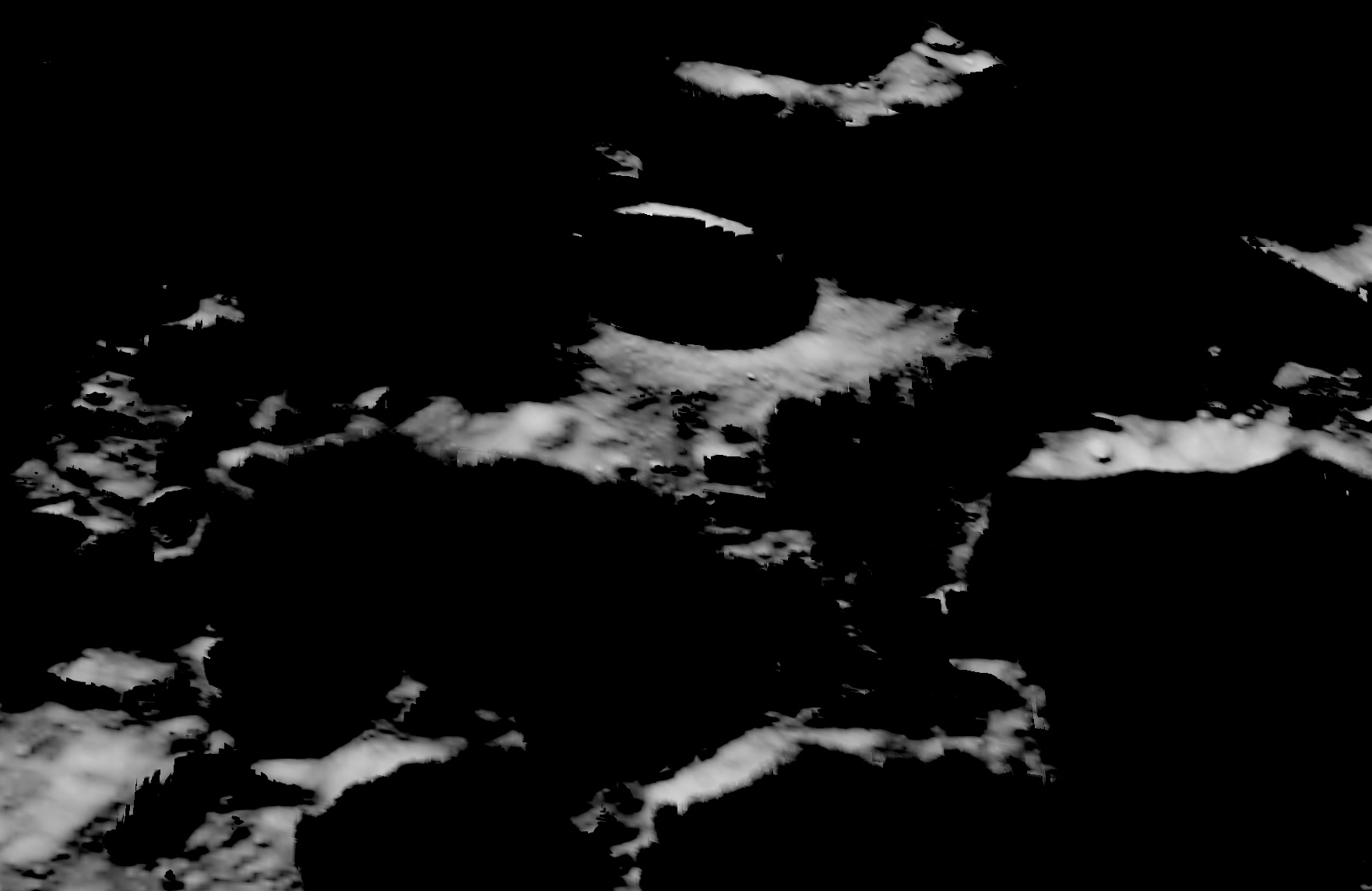}}
		\put(0.05,.8){\color{white}{(c)}}
		\end{picture}\phantomcaption{\label{subfig:nostreaks-shackleton}}
	\end{subfigure}
	\hspace{0.05\unitlength}
	\begin{subfigure}{\unitlength}
		\begin{picture}(1,1)
		\put(0,0){\includegraphics[trim=17cm 9cm 14cm 3cm, clip=true,height=\textwidth]{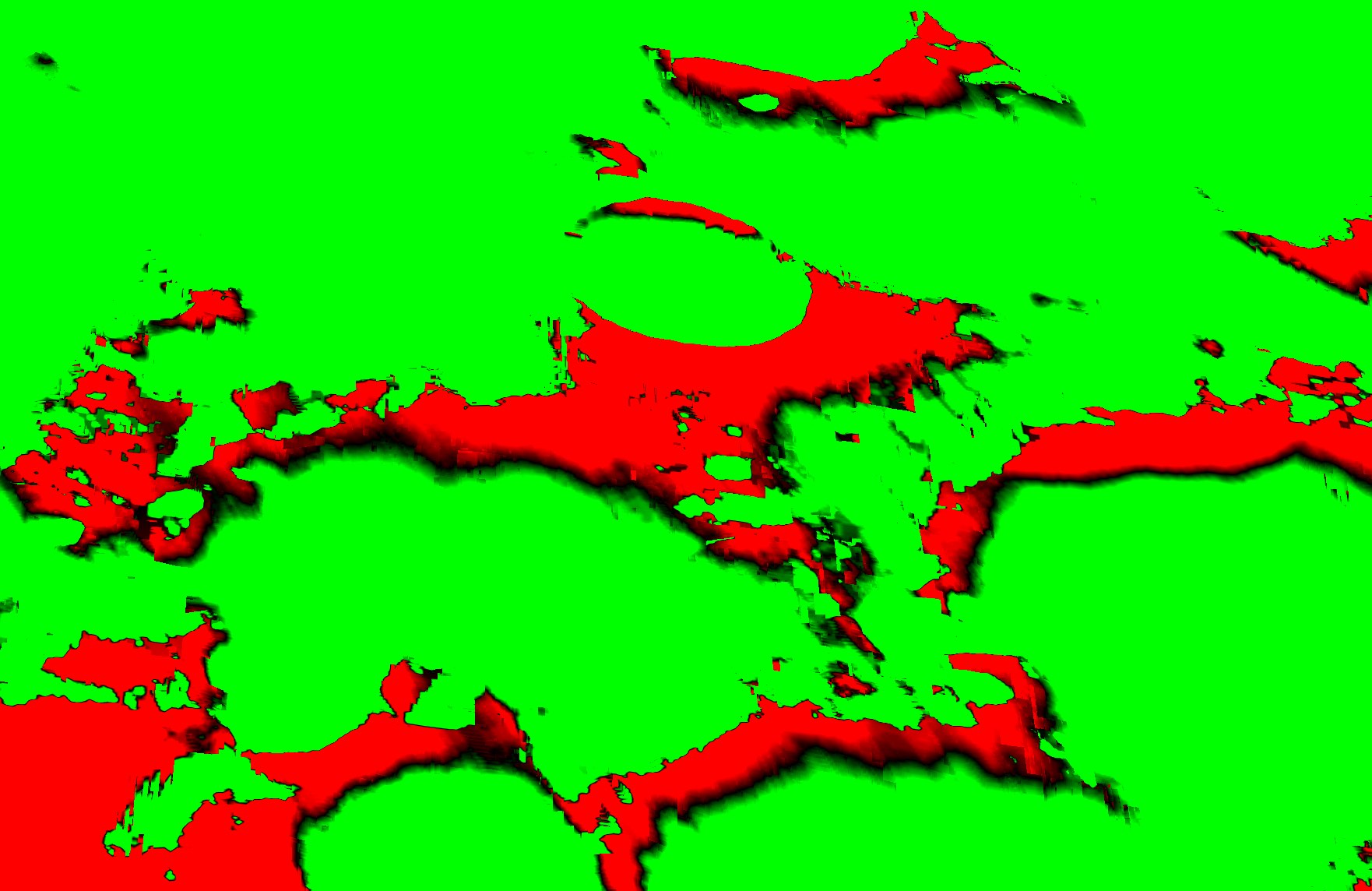}}
		\put(0.05,.8){(d)}
		\end{picture}\phantomcaption{\label{subfig:nostreaks-J}}
	\end{subfigure}
	\hspace{0.07\unitlength}
  	\begin{subfigure}{0.3333\unitlength}
	\begin{picture}(.3333,1)
		\put(0,0){\includegraphics[height=\unitlength]{cb-streaks}}
		\end{picture}
	\end{subfigure}
	\caption{Streaks when using canonical DDA (top row) vs our method (bottom).  (\subref*{subfig:streaks-J}), (\subref*{subfig:nostreaks-J}) display $J\text{*}$ cost values.}\label{fig:streakshackleton}
\end{figure}

Then, assuming that the height field curvature is nearly flat at scales of ${\sim}T$, we use a variant of DDA \cite{Amanatides1987} to traverse the maximum mipmap texels that could be intersected by the ray. This variant differs from the `canonical' DDA in that \emph{all} texels underneath the footprint of the ray are visited. \autoref{fig:streakdda} compares the two different versions of DDA, and \autoref{fig:streakshackleton} illustrates the artifacts that can result from using canonical DDA.

As \autoref{subfig:dda} shows, there are up to three intersecting texels in the interval $[0, \Delta M/2]$; we find the one that results in $\min\Delta\max h(t_{k+1})$.

The complete algorithm is given in \autoref{alg:shadowalgo}.

\begin{algorithm}[htb]
	\caption{Fast Soft Shadow Algorithm}
	\label{alg:shadowalgo}
	\SetKwInOut{Input}{Input}\SetKwInOut{Output}{Output}%
	\DontPrintSemicolon%
	\SetAlgoNoEnd%
	$J\raisebox{0.1em}{\text{*}}_{\hspace{-.5em}0}\leftarrow1$\;
	$t_0\leftarrow 1$\;
	$m\leftarrow N'-1$\;
	$\Delta\mathbf{R}\leftarrow$Texel step size\;
	\For{$k\leftarrow 0$ \KwTo $N'-1$}{
		$t\text{*}\leftarrow -1$, $\Delta\max h\text{*}\leftarrow1$, $i\leftarrow0$\;
		Compute height $H$ of ray tip at $R(t_{k,i})$\;
		Sample $\max h(t_{k+1})$ at $R(t_{k,i})$, mip level $m$\;
		$\Delta\max h_k\leftarrow H - \max h(t_{k+1})$\;
		\If{$\Delta\max h_k < \Delta\max h\text{*}$}{
			$\Delta\max h\text{*}\leftarrow\Delta\max h(t_{k+1})$\;
			$t\text{*}\leftarrow t_k$\;
		}
		\For(\tcp*[f]{DDA}){$i\leftarrow 1$ \KwTo $2$}{
			$R(t_{k,i})\leftarrow R(t_{k,i-1}) - 2^{-k-1}\Delta\mathbf{R}$\;
			$t_k\leftarrow t_k - 2^{-k-1}$\;
			Compute height $H$ of ray tip at $R(t_{k,i})$\;
			Sample $\max h(t_{k+1})$ at $R(t_{k,i})$, level $m$\;
			$\Delta\max h_k\leftarrow H - \max h(t_{k+1})$\;
			\If{$\Delta\max h_k < \Delta\max h\text{*}$}{
				$\Delta\max h\text{*}\leftarrow\Delta\max h_k$\;
				$t\text{*}\leftarrow t_k$\;
			}
		}
		$m\leftarrow m - 1$, $k\leftarrow k+1$\;
		\If{$t\text{*} > -1$}{
			$t_k\leftarrow t\text{*}+2^{-k}$\;
		}
	}
	\If{$\Delta\max h\text{*} < 1$}{
		$J\text{*}\leftarrow(\Delta\max h\text{*})/t\text{*}$\;
	}
\end{algorithm}

\subsection{Scalable Rendering}
\label{subsec:scalablerender}

Next we discuss several important additional details for scalable rendering including virtual texture coordinates, and interval concatenation for handling long shadows.

\noindent\textbf{Virtual texture mapping}
For our renderer to scale from planetary radii of thousands of kilometers to surface features of tens of meters (a scale factor of ${\sim}10^5$) with the limited VRAM of a commodity graphics card, we employ a tile-based virtual texture scheme and stitch together a subset of visible tiles at runtime. While allowing us to render scenes across a wide range of distance scales, this also implies that care must be taken to use scaled and offset texture coordinates of the stitched texture when sampling.

\noindent\textbf{Long shadows} We noted in \autoref{subsec:dynamicprog} that we could use less mip levels than $\log N$ as shadows are not usually very long in practice. But this also has a beneficial effect of mitigating artifacts when approximating $\max h(t_{k+1})$ with a maximum mipmap. This is because the effect of an error is amplified at higher mip levels $k\rightarrow0$ due to subdivision interval sizes doubling at each level: at mip level 0 for instance, $t=\{0,\frac{1}{2},1\}$ and supposing that $t=1$ was chosen instead of $t=0$ subsequent steps will always be offset by 1, while if the algorithm is started at mip level 5 then the initial error in $t$ will be $\leq2^{-5}$.

However the chosen initial mip level $N'$ may not be sufficiently large to cover very long shadows. 
Instead we run \autoref{alg:shadowalgo} again, but setting the sampling interval immediately after the first interval, i.e., $t_{N'+k}=t_0+T+t_k$. As typically the total number of required iterations $N'$ need not be very large, this can be repeated once or twice as needed. $N'=\{5,5,5\}$ for all scenes in this work unless specified.


\begin{figure}[htb]
    \centering
    \begin{subfigure}{0.4\textwidth}
        \includegraphics[width=\textwidth]{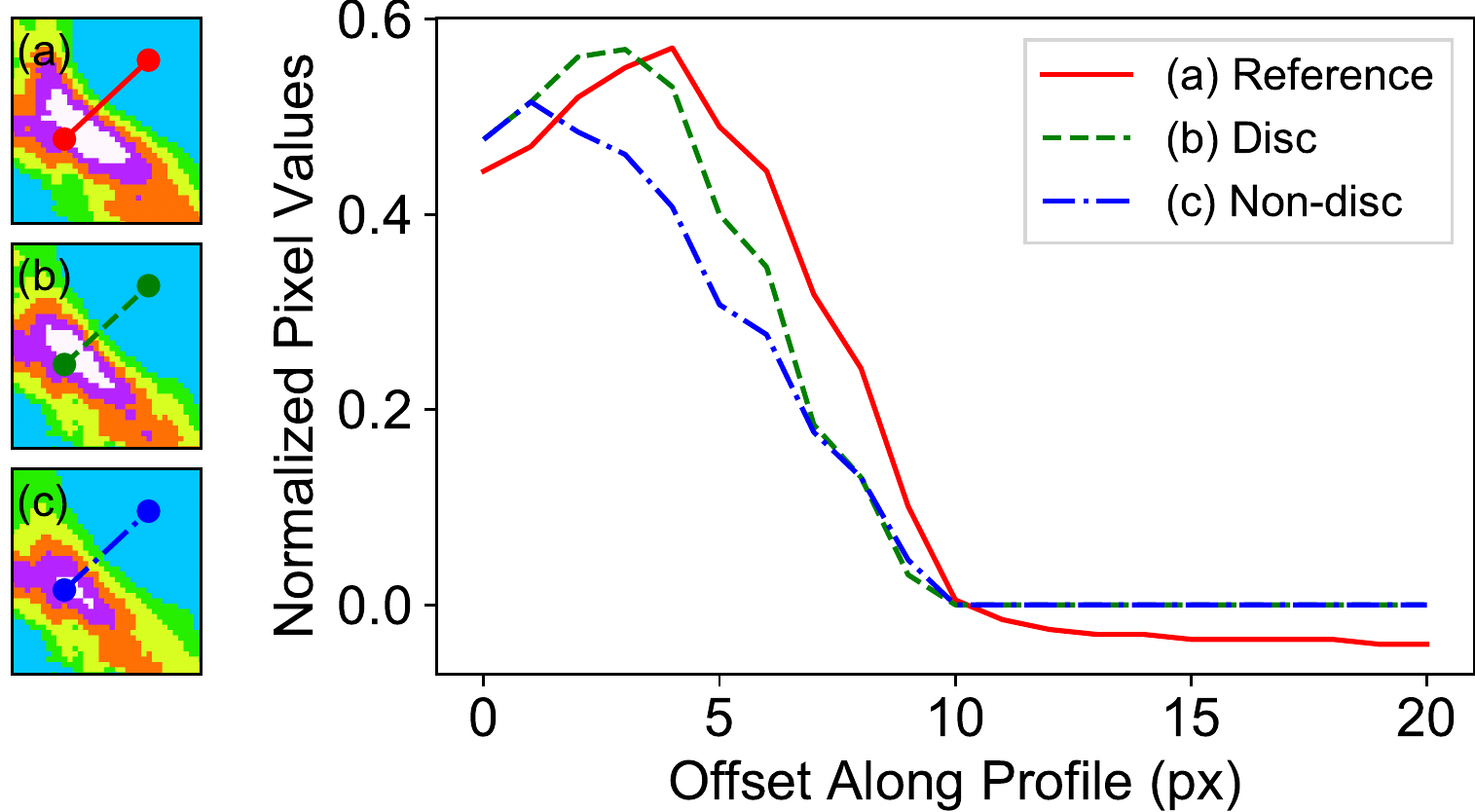}
    \end{subfigure}\caption{Normalized pixel value profiles (scaled between the minimum and maximum value ranges for each image to highlight slope differences) across lit-shadow boundary for (a) LRO WAC reference image\cite{Speyerer2013}, (b) our render using disc occlusion and (c) without. (b) tracks the reference profile more closely, while (c) underestimates values near the left side of the profile. \label{fig:discocclusion}}
\end{figure}

\subsection{Sun Occlusion}
\label{subsec:sunocclusion}
We implemented a physically-based penumbra calculation that assumes light sources (e.g., Sun, Earth) are circular discs emitting a cone of constant light flux with the apex at the view ray intersection point. Moreover, we make the simplifying assumption that the cross section of the terrain shadowing the light is a prism, i.e., the terrain height is constant in the direction perpendicular to the shadow ray. This allows us to calculate the shadow fraction $s$ as a circular segment. (\autoref{subfig:penumbrageom}). While real light sources such as the Sun exhibit limb darkening, and cross sections of terrain are usually never prisms, nevertheless our disc occlusion method generates more accurate results over other \cite{Tatarchuk2006,Tamas2016}, non-disc based approximations. \autoref{fig:discocclusion} compares disc occlusion and non-disc methods with a calibrated ground-truth image from the Lunar Reconnaissance Orbiter (LRO) Wide Angle Camera (WAC)\cite{Speyerer2013}.

Defining $r_L$ as the apparent light radius, surface normal $\mathbf{\hat{N}}$, normal vector aligned to be perpendicular to the shadow ray $\mathbf{\hat{N}}_{\hat{L}}$, the distance $d$ from the center of the light disc to the edge of the shadowed segment, normalized to disc radius=1, is:
\begin{align}\label{eq:discsegment}
d &= 2\frac{J\text{*}(\mathbf{\hat{N}}\cdot\mathbf{\hat{N}}_{\hat{L}})-r_L}{r_L}&&\begin{aligned}d\in[-1, 1]\\
J\text{*} < 1
\end{aligned}
\end{align}

The shadow fraction $s$ is then:
\begin{equation}\label{eq:discfrac}
s = \max\left\{\frac{\pi-\cos^{-1}d+d\sqrt{1-d^2}}{\pi}, 1-J\text{*}\right\}
\end{equation}

$\mathbf{\hat{N}}\cdot\mathbf{\hat{N}}_{\hat{L}}$ can be computed in the vertex shader for speed.

\section{View Ray Casting}
\label{sec:viewraycasting}

\begin{figure}[htb]
	\centering
	\includegraphics[width=0.3\textwidth]{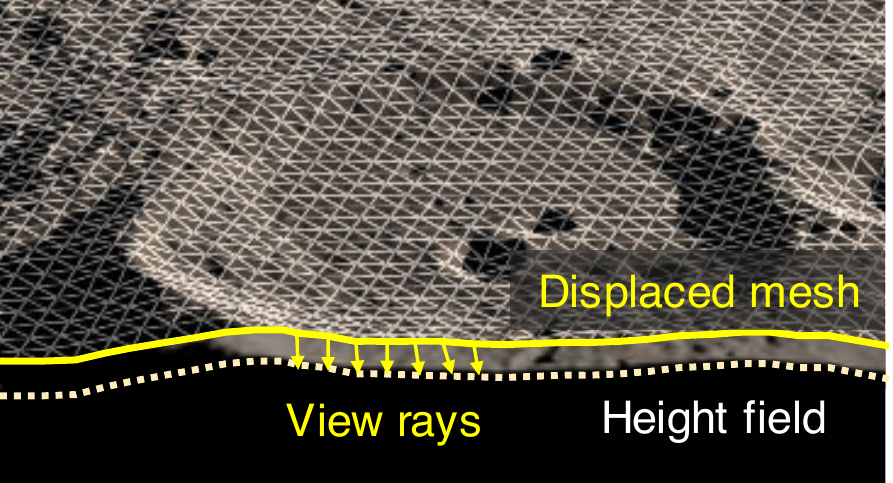}
	\caption{Hybrid ray casting scheme. Increased accuracy is obtained by shooting rays from vertices pre-displaced with the height field used to determine ray visibility.}
	\label{fig:raydiagram}
\end{figure}

View ray casting can result in artifacts due to inaccurate ray-surface intersections. To combat this, we developed a novel hybrid ray casting approach (\autoref{fig:raydiagram}), where we start from a cubesphere geometry surface, which is a spherical mesh that entirely contains the lunar terrain at the maximum terrain height, and first displace vertices inward in the vertex shader with the same height field used to determine ray visibility. Then rays are shot from the displaced vertices $\mathbf{p}$ in the fragment shader.

Starting from pre-displaced vertices has a couple of advantages. The biggest is the decreased number of steps required, because rays are already starting close to their final intersection points.

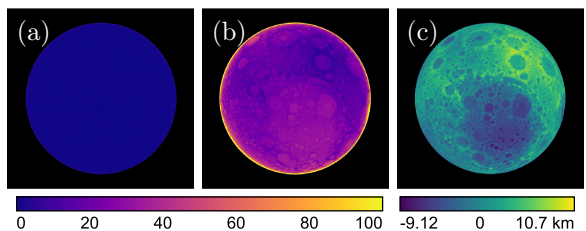
\begin{figure}[tbp]
    \centering
    \begin{subfigure}{0.15\textwidth}
        \setlength{\unitlength}{1.0\textwidth}
        \begin{picture}(1,1)
        \put(0,0){\includegraphics[trim=11cm 0cm 11cm 0cm, clip=true, width=\textwidth]{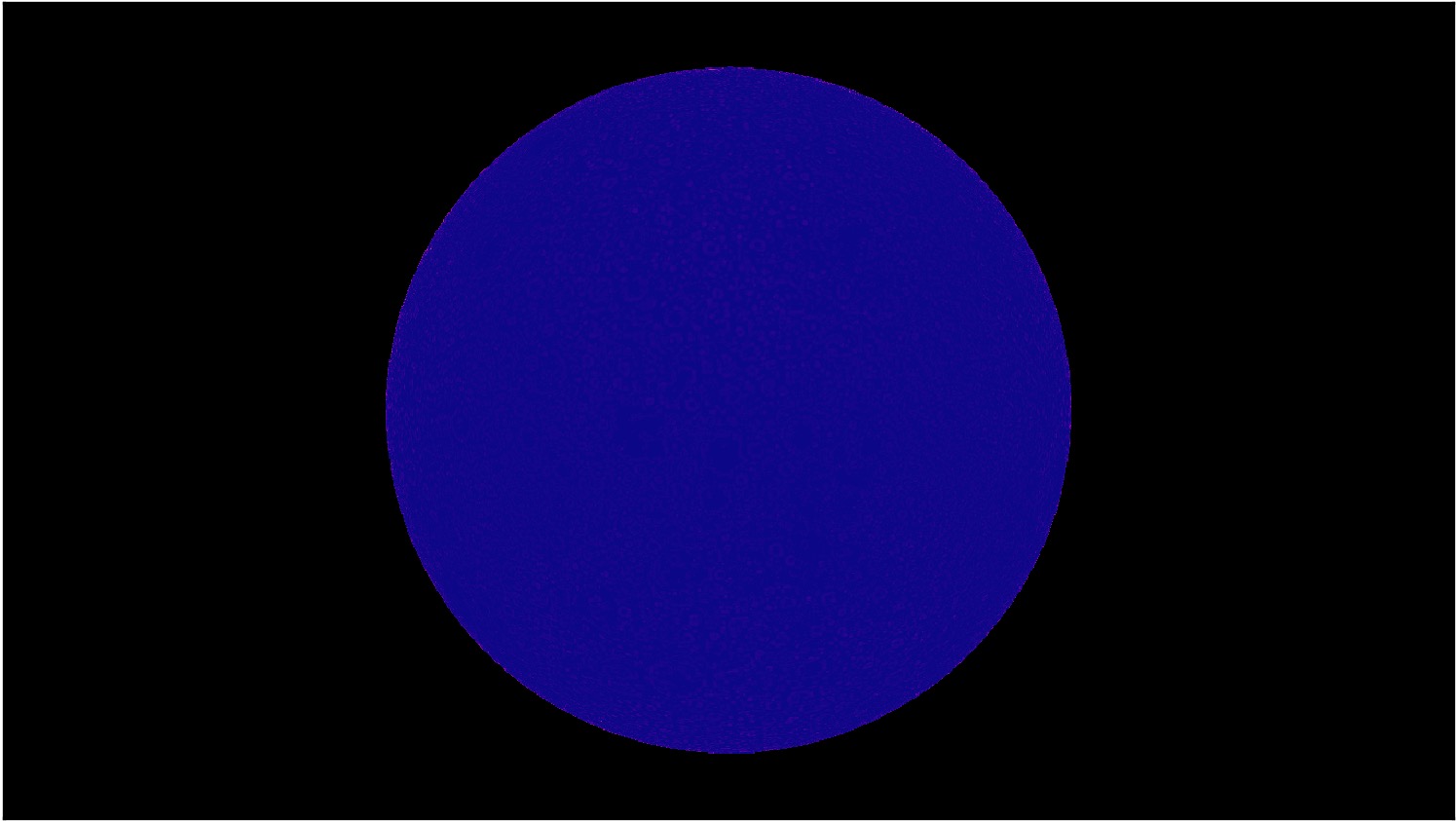}}
        \put(0.05,0.8){\color{white}{(a)}}
        \end{picture}\phantomcaption{\label{subfig:hybrid}}
    \end{subfigure}
    \begin{subfigure}{0.15\textwidth}
        \setlength{\unitlength}{1.0\textwidth}
        \begin{picture}(1,1)
        \put(0,0){\includegraphics[trim=11cm 0cm 11cm 0cm, clip=true, width=\textwidth]{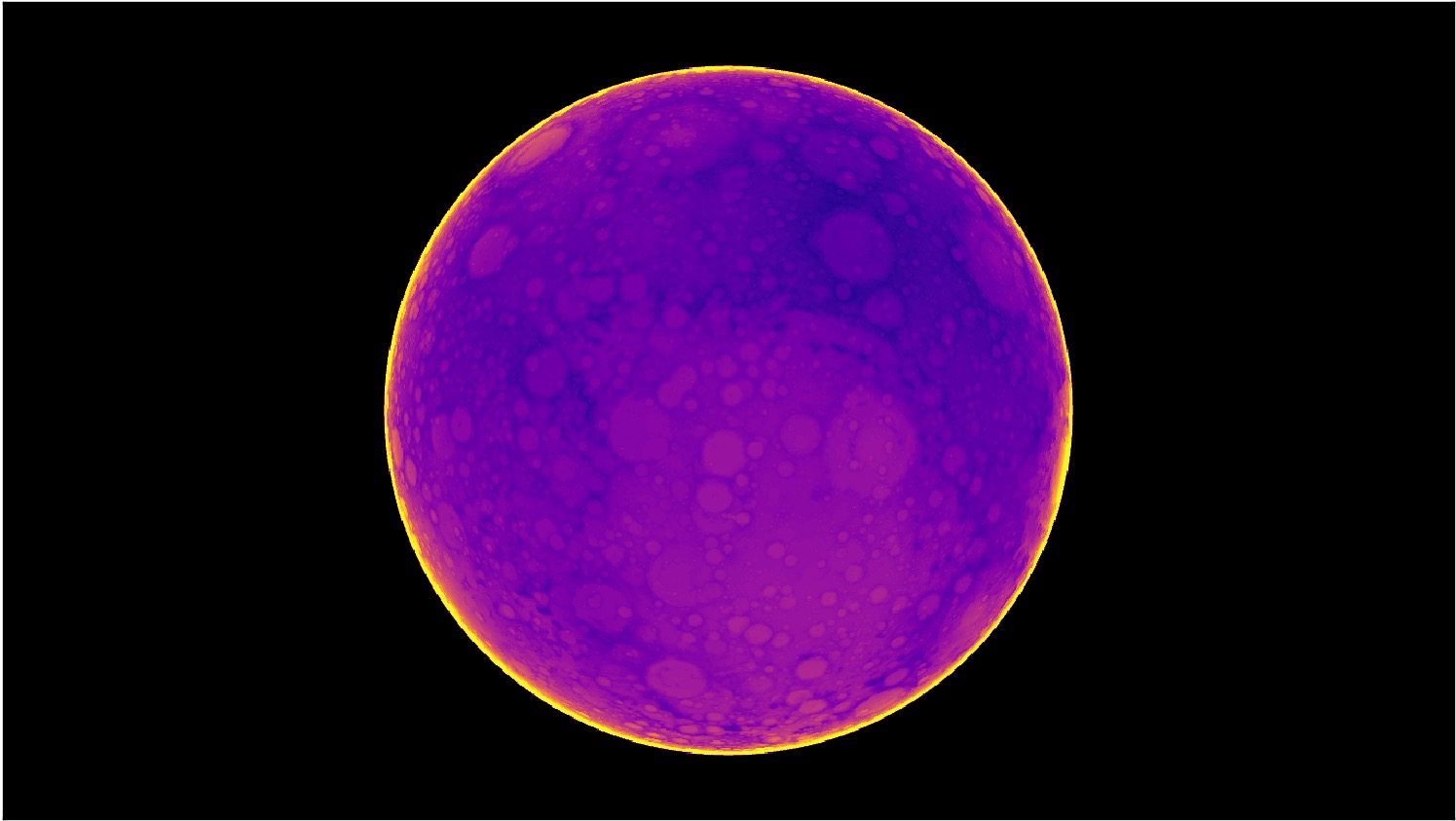}}
        \put(0.05,0.8){\color{white}{(b)}}
        \end{picture}\phantomcaption{\label{subfig:nodisp}}
    \end{subfigure}
    \begin{subfigure}{0.15\textwidth}
        \setlength{\unitlength}{1.0\textwidth}
        \begin{picture}(1,1)
        \put(0,0){\includegraphics[trim=11cm 0cm 11cm 0cm, clip=true, width=\textwidth]{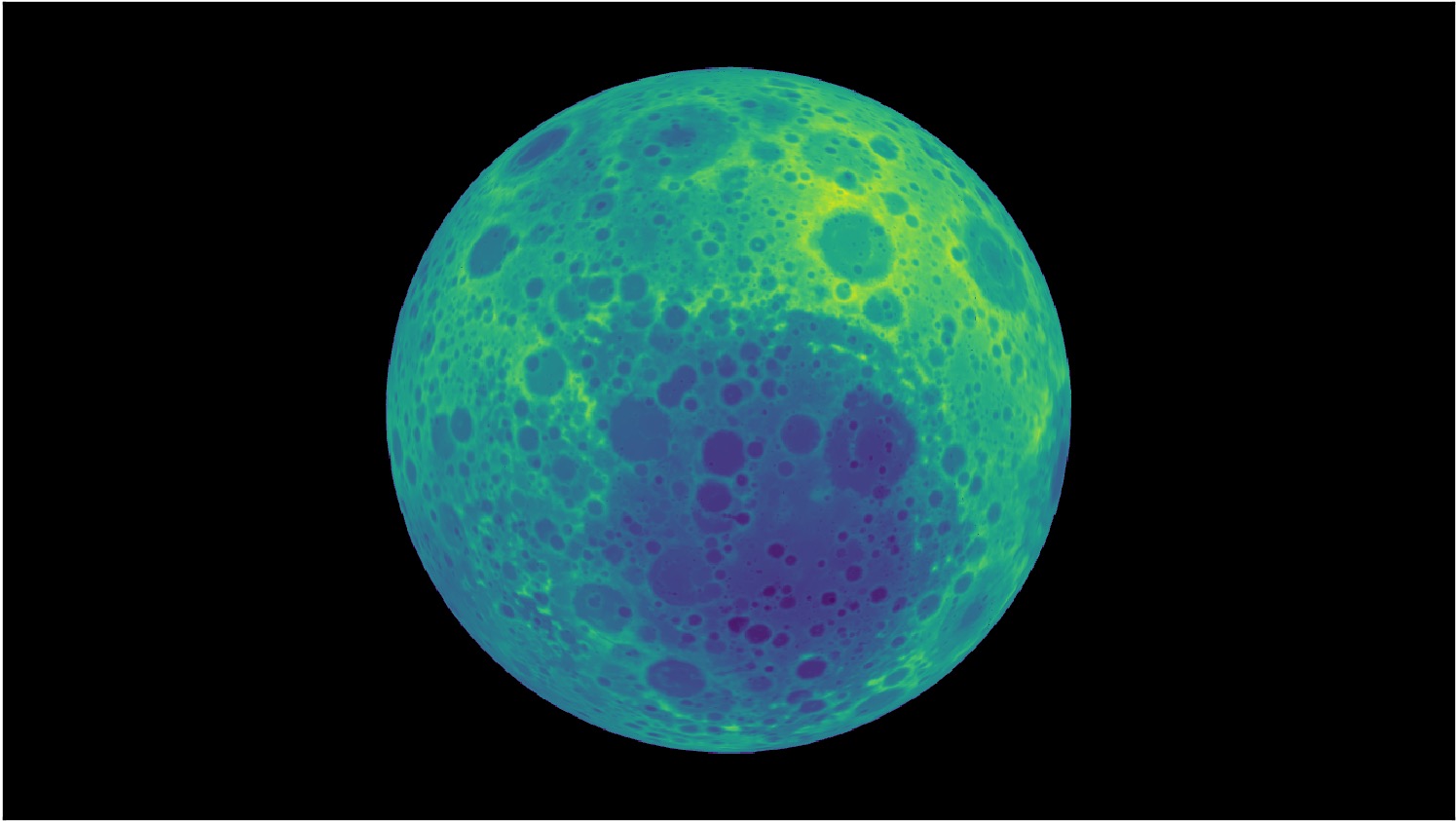}}
        \put(0.05,0.8){\color{white}{(c)}}
        \end{picture}\phantomcaption{\label{subfig:dispheight}}
    \end{subfigure}
    \\
    \begin{subfigure}{0.45\textwidth}
        \includegraphics[width=\textwidth]{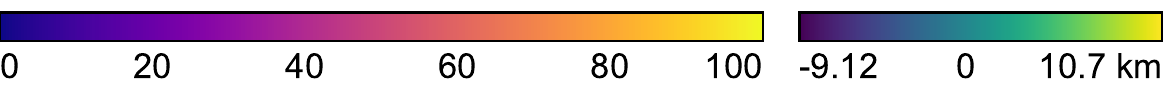}
    \end{subfigure}%
    \caption{View ray step counts with vertex pre-displacement (\subref*{subfig:hybrid}), and (\subref*{subfig:nodisp}) without. Altitudes are shown in (\subref*{subfig:dispheight}).
Without pre-displacement, nearly 50--100 times more steps are required to trace the planet limb and low altitudes.\label{fig:viewstepvis}}
\end{figure}

\begin{figure}[tb]
    \centering
    \setlength{\unitlength}{1.0\textwidth}
    \begin{picture}(0.4,0.2)
    \put(0,0){\includegraphics[width=0.4\textwidth]{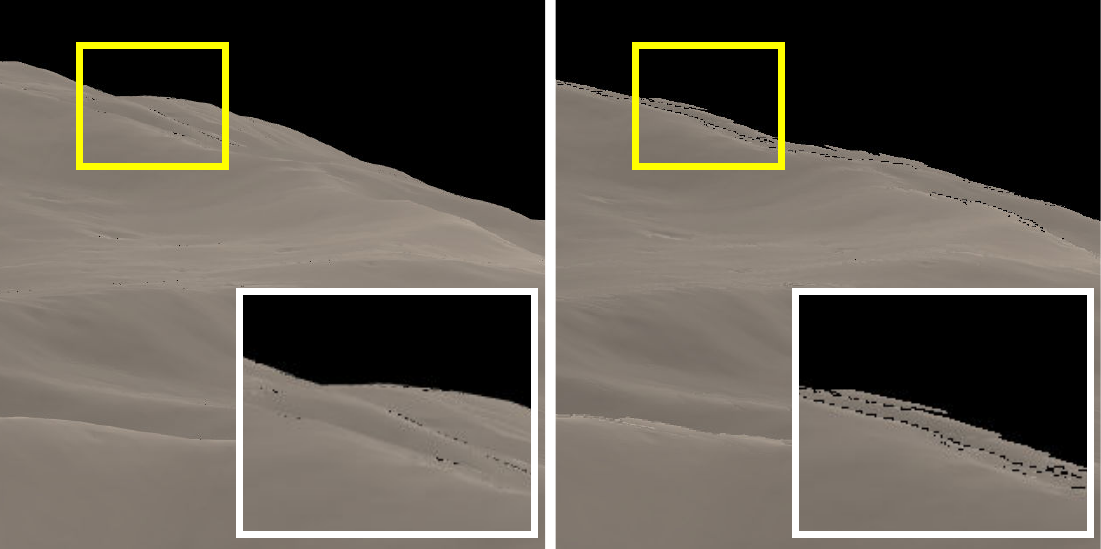}}
    \put(0.17,0.177){\color{white}{(a)}}
    \put(0.37,0.177){\color{white}{(b)}}
    \end{picture}
    \begin{subfigure}{0\textwidth}%
        \phantomcaption{\label{subfig:vdispafter}}
    \end{subfigure}
    \begin{subfigure}{0\textwidth}%
        \phantomcaption{\label{subfig:vdispbefore}}
    \end{subfigure}
    \caption{With vertex pre-displacement (\subref*{subfig:vdispafter}) and without (\subref*{subfig:vdispbefore}). Ray casting without vertex pre-displacement results in ugly artifacts and missed terrain, problems which are gone in our hybrid method.}\label{fig:vertexdisp}
\end{figure}

As \autoref{fig:viewstepvis} shows, ray casting without pre-displacement requires up to 100 times the number of steps to reach an equivalent level of accuracy and can still result in artifacts at the object silhouette (\autoref{fig:vertexdisp}).
Mesh resolution does not need to be very high (we use a cube sphere with up to $4096$ subdivisions per cube face depending on the lod), as the purpose of displacement is simply to bootstrap ray starting points to positions near the actual surface.

Aliasing artifacts occur due to point sampling, and also because of incorrect automatic mip level deduction caused by discontinuous sampling of textures.
We were able to reduce temporal flickering by reducing sample rates (view ray step size) at highly oblique angles where $\mathbf{\hat{N}}\cdot\mathbf{\hat{V}}\rightarrow0$, in a form of low-pass filtering. We mitigated the second issue by explicitly calculating texture gradients using the method of tracing ray differentials \cite{Igehy1999}.

\begin{figure}[hb]
    \centering
    \includegraphics[width=0.45\textwidth]{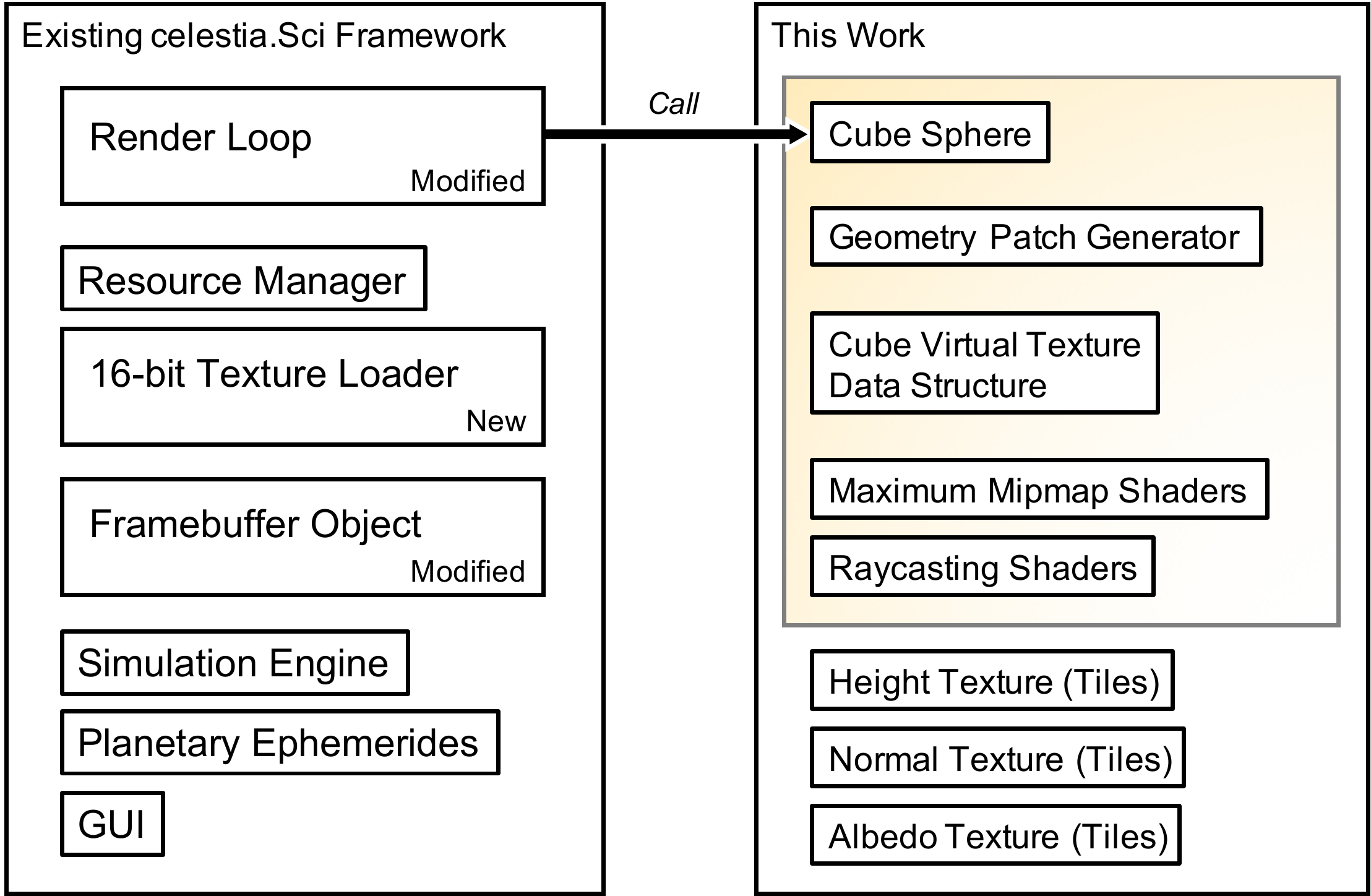}%
    \caption{Our work within celestia.Sci\label{fig:celestiablock}}
\end{figure}

\section{celestia.Sci}
\label{sec:celestia}

celestia.Sci \cite{Schrempp2013Welcome} 
is a real-time, three-dimensional,
interactive space simulation software that can render objects at a large range
of scales, from Solar System objects to deep space and galaxies
\cite{Schrempp2013Welcome}.
The software is based on Celestia
\cite{Laurel2001}. Celestia has been used by NASA \cite{NASA2010} and ESA
\cite{Esa2004Closing} due to its visualization accuracy and extensive
astronomical database built on peer-reviewed scientific data.

celestia.Sci provides a scene management and framebuffer display framework, support for reading solar system and spacecraft ephemerides, and
scripting and interactive features. This work adds shaders and new \texttt{height} and \texttt{albedo} texture types that can optionally be used in an addon (a directory defining a new or modified celestial body). These new texture types consist of tiles organized in cube face directories and their use in an addon automatically activates the new shaders. This design adds new features to celestia.Sci while modifying existing code as little as possible. \autoref{fig:celestiablock} outlines how our work integrates with celestia.Sci.

\begin{figure*}[t]
	\centering
	\begin{subfigure}{0.3\textwidth}
		\includegraphics[width=\textwidth]{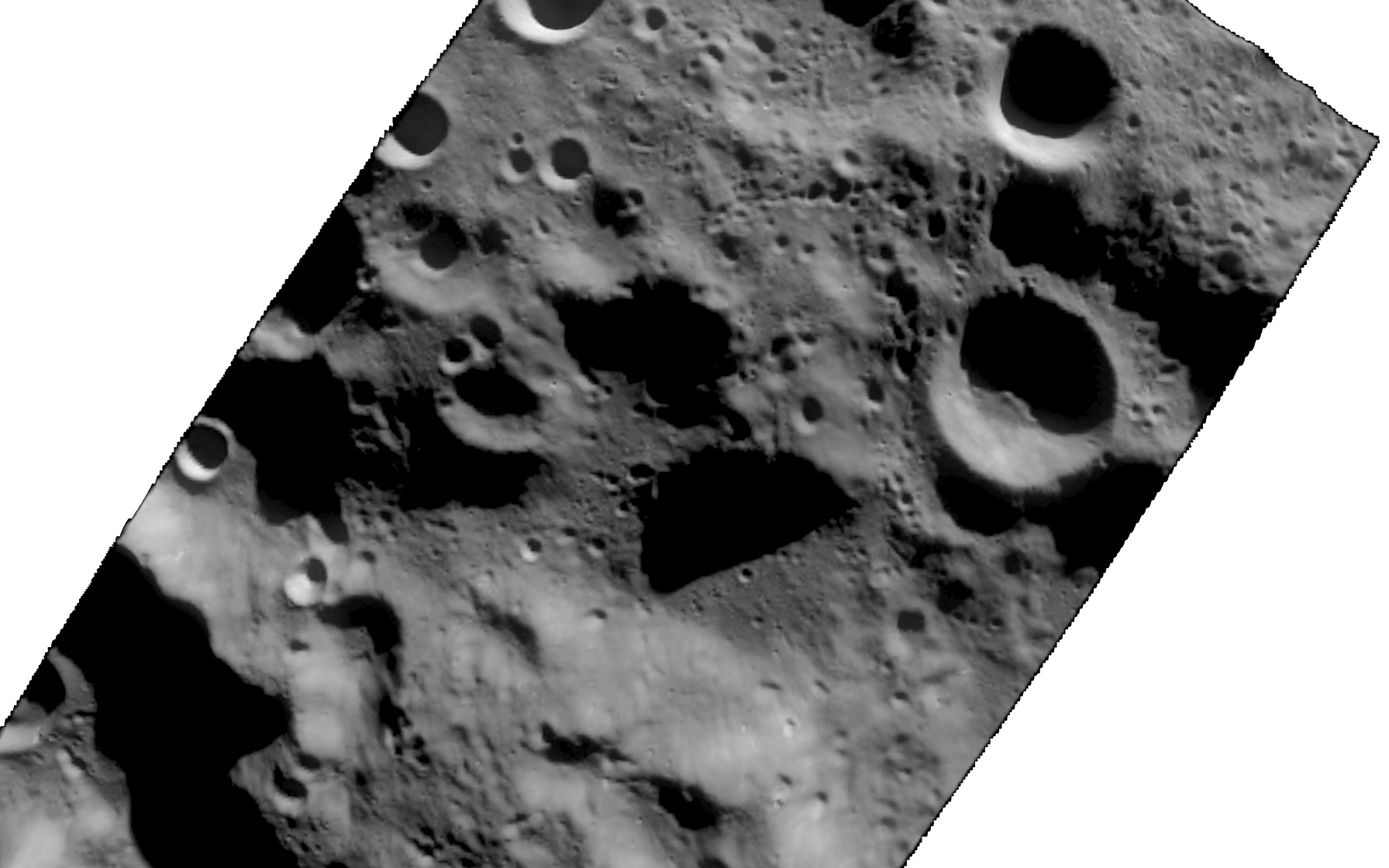}
		\caption{\label{subfig:wacpolar}}
	\end{subfigure}
	\begin{subfigure}{0.3\textwidth}
		\includegraphics[width=\textwidth]{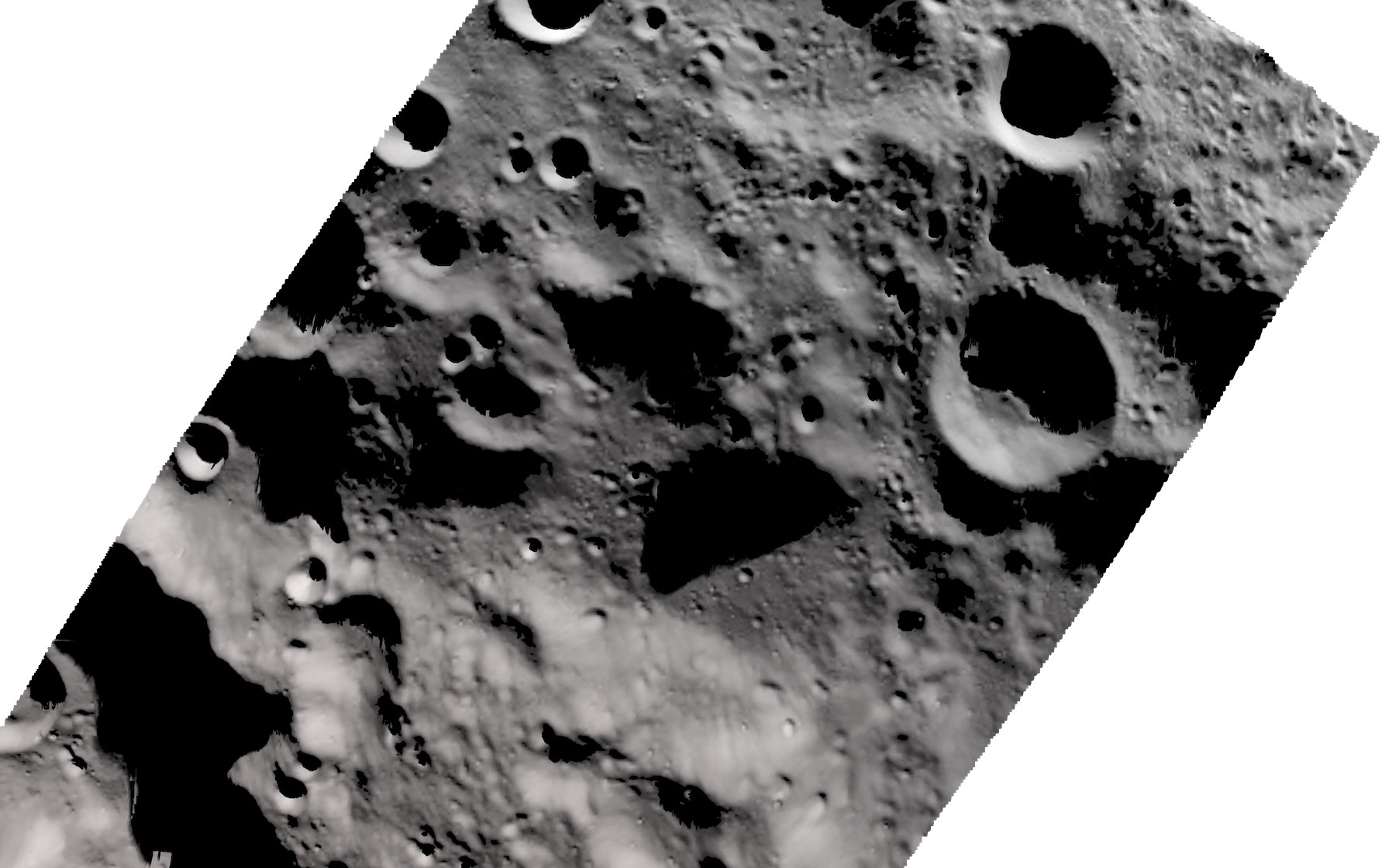}\caption{\label{subfig:wacrender}}
	\end{subfigure}
	\begin{subfigure}{0.34\textwidth}
		\includegraphics[width=\textwidth]{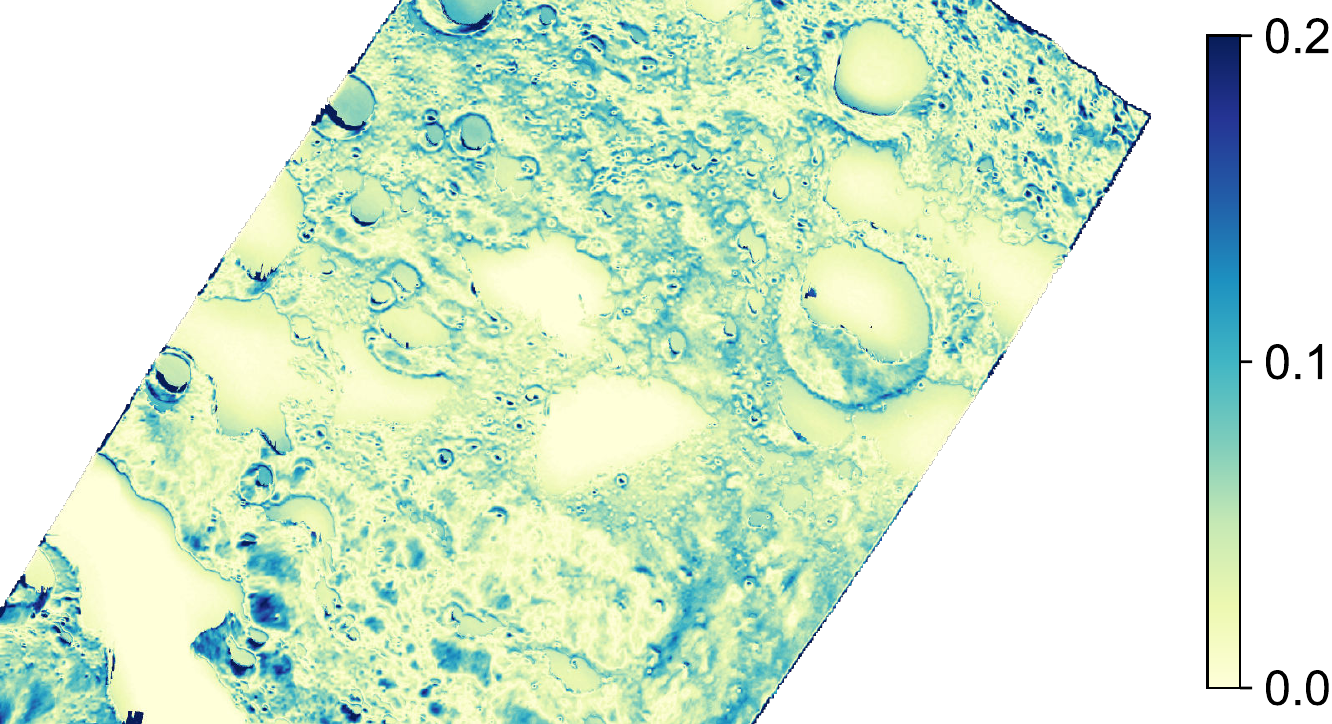}\caption{\label{subfig:wacerror}}
	\end{subfigure}
	\caption{LRO WAC polar imagery (\subref*{subfig:wacpolar}), Our render (\subref*{subfig:wacrender}), Error (\subref*{subfig:wacerror})}\label{fig:resultspolar}
\end{figure*}

\section{Results}
\label{sec:results}

Here we present results comparing our simulations to actual Moon imagery.

\subsection{South Pole Simulation}
\label{subsec:results-wac}

We compared south pole LRO WAC imagery with our simulations. As our renderer does not produce georeferenced images, we instead render the WAC images unshaded in celestia.Sci by draping the images onto the same Moon terrain mesh used by our ray casting engine (we use double the mesh frequency to compensate for the comparatively low resolution). This guarantees that our view of the WAC images exactly register with our simulations. Moreover a Hapke BRDF \cite{Sato2014} and earthshine (indirect light from the Earth) was used to closely simulate lunar surface lighting.


WAC images are \textit{radiance} while our simulations produce \textit{irradiance} values; fortunately as these only differ by a constant multiplicative factor we perform a relative comparison of normalized images by dividing all pixel intensities by their maximum.  

An example set of results is shown in \autoref{fig:resultspolar}. For this 
case, the standard deviation of the error $\sigma\approx 0.040$, and 97\% of pixels are within $3\sigma$.
Some of the `errors' are in fact due to a diffuse glow surrounding bright pixels in the WAC images that bleeds into adjacent shadows, brightening them. This is a known artifact called \textit{ghosting} that is caused by stray light in the camera \cite{Mahanti2016} and which we do not try to simulate. To compensate, we subtract empirically determined dark pixel values from the normalized WAC images before computing errors. For \autoref{fig:resultspolar} we subtract a dark value of 7/255; additional examples in \autoref{fig:morepolarmovie} subtract 5/255. We attribute other small errors near shadow boundaries mainly to artifacts from using a maximum mipmap (see \autoref{subsec:dynamicprog}) that have not been completely mitigated by the technique described in \autoref{subsec:scalablerender}.


\begin{figure*}[htbp]
	\centering
	\begin{subfigure}{0.75\textwidth}
		\setlength{\unitlength}{1.0\textwidth}
		\begin{picture}(1,0.55)
		\put(0,0){\includegraphics[width=\textwidth]{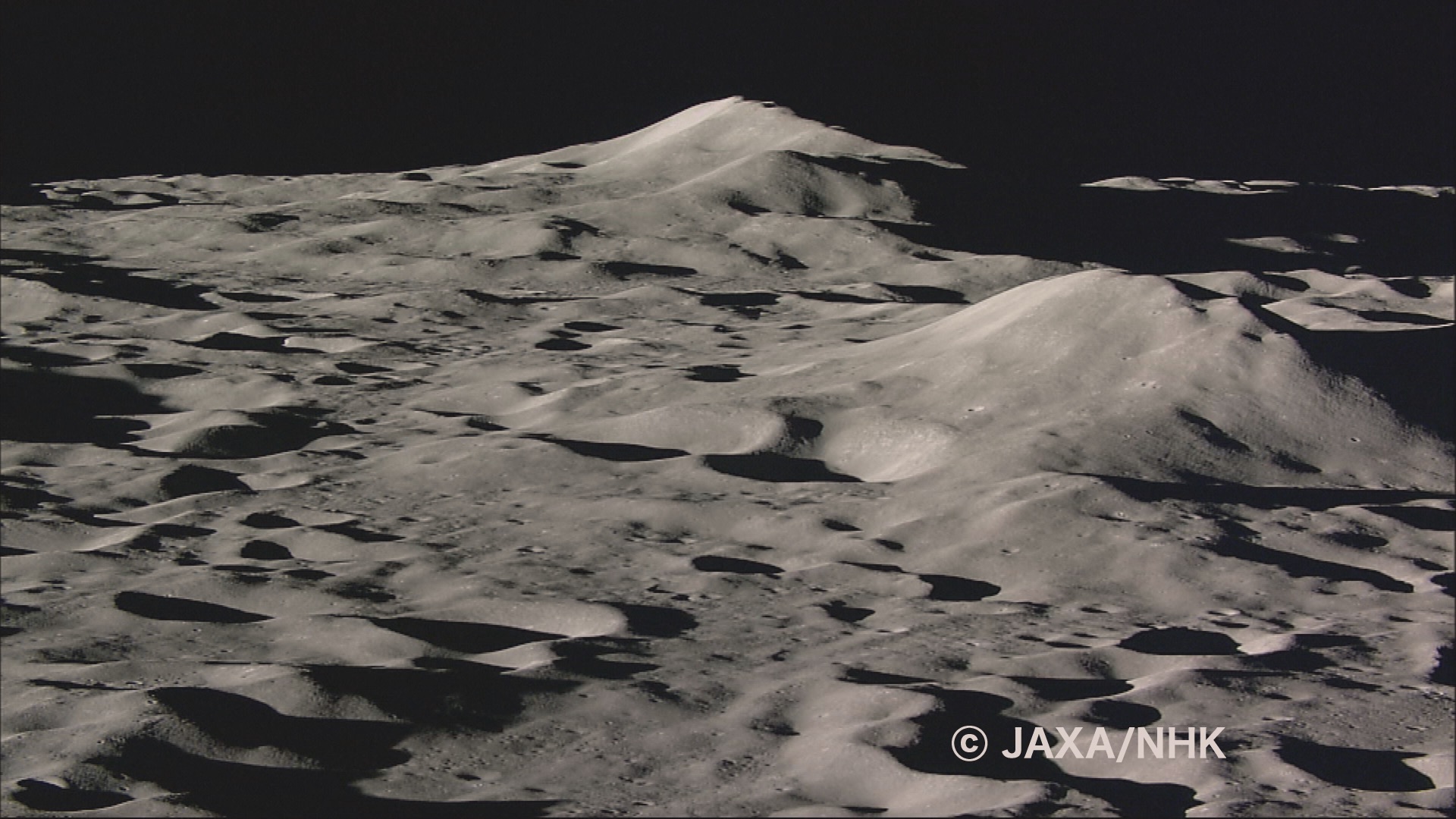}}
		\put(0.023,0.51){\color{white}{(a)}}
		\end{picture}\phantomcaption{\label{subfig:hdtv}}
	\end{subfigure}\\
	 \vspace{1em}
	\begin{subfigure}{0.75\textwidth}
		\setlength{\unitlength}{1.0\textwidth}
		\begin{picture}(1,0.55)
		\put(0,0){\includegraphics[trim=0cm 6cm 0cm 2cm,clip=true,width=\textwidth]{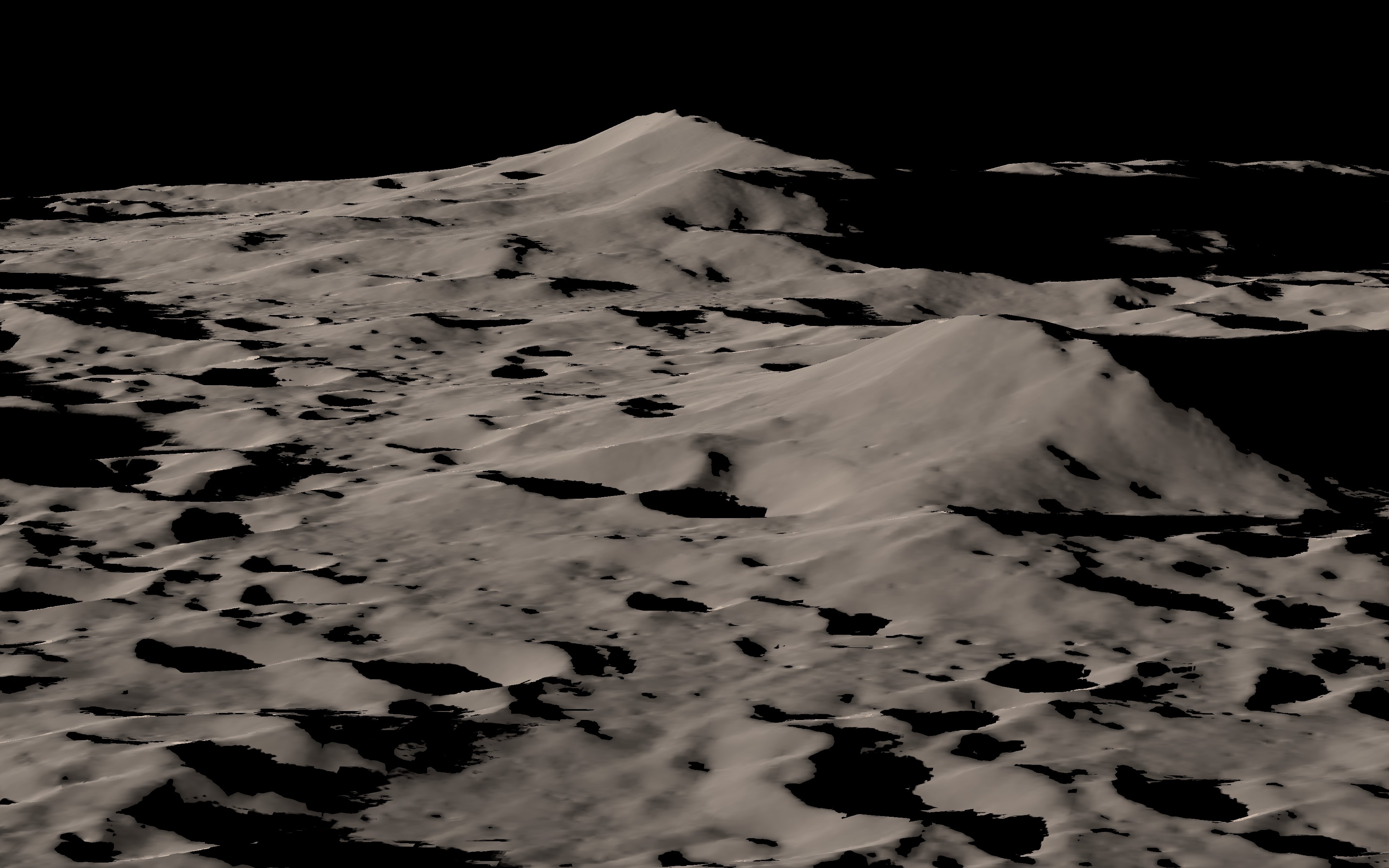}}
		\put(0.023,0.51){\color{white}{(b)}}
		\end{picture}\phantomcaption{\label{subfig:hdtvsim}}
	\end{subfigure}%
	\caption{SELENE imagery (\subref*{subfig:hdtv}), \textcopyright~JAXA/NHK, Our render (\subref*{subfig:hdtvsim}).}
	\label{fig:resultskaguya}
\end{figure*}

\subsection{SELENE Descent Comparison}
\label{subsec:results-selene}

We compare HDTV frames from the SELENE mission \cite{Yamazaki2010} with our simulation. The frames \cite{JAXA} were captured during SELENE's final collision trajectory near the lunar south pole and are thus a useful analogue for images captured during a polar landing. 
Following Honda et al.~\cite{Honda2010}, we color-calibrate the raw HDTV frames except we omit dark offset subtraction as no reliable values could be found that did not create an unnatural color cast. We also applied irradiance scale factors $C_{rr}$ from Honda et al. as values normalized to $[0, 1]$ and finally applied the same sRGB transfer (gamma) function used in celestia.Sci for a consistent comparison. The same Hapke BRDF used in \autoref{subsec:results-wac} was used here.
Results are shown in \autoref{fig:resultskaguya}.
 The distinctive mottled appearance of shadows and pale pinkish brown color of the terrain is closely matched by our render.

\subsection{Performance}
\label{subsec:results-perf}

\begin{figure*}[t]
    \centering
            \begin{overpic}[width=0.82\textwidth]{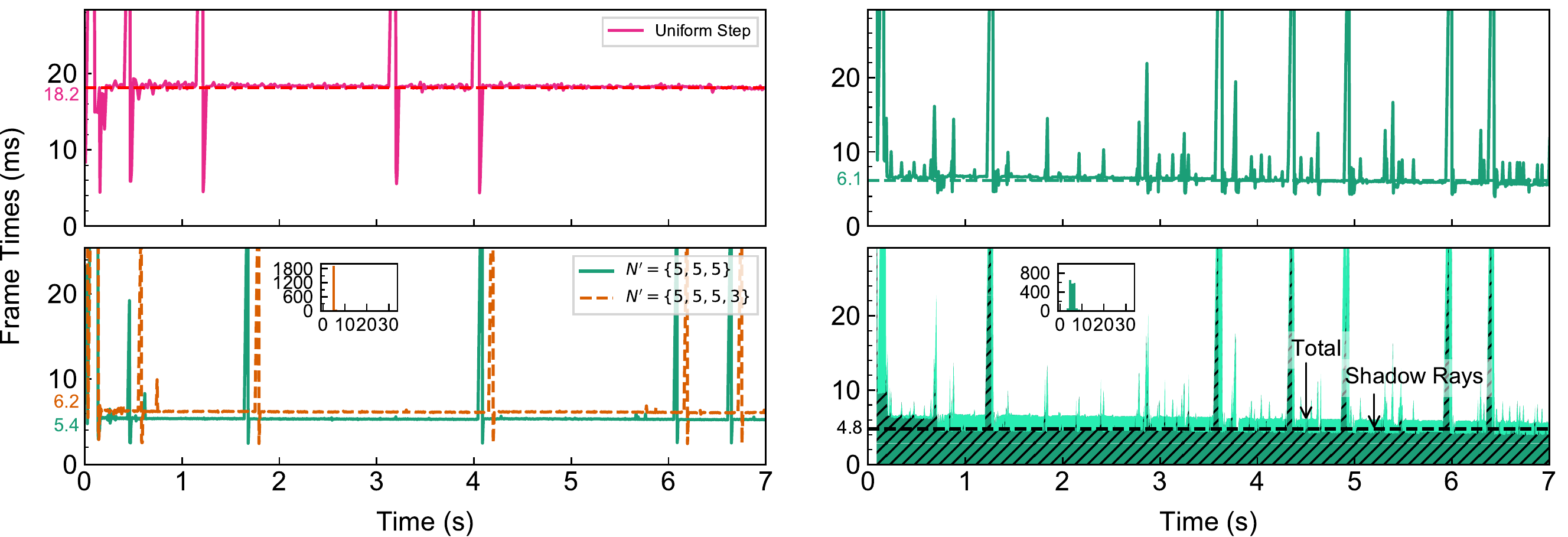}
                \put(45.8,20.5){\setlength{\fboxsep}{0.5pt}\setlength{\fboxrule}{0pt}\fcolorbox{white}{white}{(a)}}\phantomsubcaption{\label{subfig:timen}}
                \put(45.8,5.3){\setlength{\fboxsep}{0.5pt}\setlength{\fboxrule}{0pt}\fcolorbox{white}{white}{(b)}}\phantomsubcaption{\label{subfig:timenoptim}}
                \put(95.9,20.5){\setlength{\fboxsep}{0.5pt}\setlength{\fboxrule}{0pt}\fcolorbox{white}{white}{(c)}}\phantomsubcaption{\label{subfig:timedescent}}
                \put(95.9,5.3){\setlength{\fboxsep}{0.5pt}\setlength{\fboxrule}{0pt}\fcolorbox{white}{white}{(d)}}\phantomsubcaption{\label{subfig:timebreakdown}}
            \end{overpic}
    \caption{Left: Frame times for SELENE scene near lunar surface. Histogram (inset) indicates most frame times are clustered around the mean.
		Compared to uniform stepping  (\subref*{subfig:timen}), 
        our method (\subref*{subfig:timenoptim}) is up to $237$\% faster with 185 Hz frame rates. The effect on frame times of varying $N'$ is also shown. Right: Frame times for a descent scene with >100 km vertical travel (\subref*{subfig:timedescent}), (\subref*{subfig:timebreakdown}).}%
    \label{fig:fraps-real}
\end{figure*}
Our hardware is a laptop with Intel Core i7-4900MQ CPU, 16GB RAM, and NVIDIA GTX 970M graphics with 6GB VRAM (%
we disabled Scalable Link Interface (SLI)). We ran our performance tests on the laptop's SATA III SSD rated at 6Gb/s and the 1920$\times$1080 internal display with vsync and MSAA off. 


We tested the scene shown in \autoref{fig:resultskaguya} with the SELENE spacecraft passing over two mountains, and also a fictitious descent trajectory covering >100 km vertical distance. The latter is a test of the ability of our system to smoothly handle a wide range of distance scales. Stitched height fields and maximum mipmaps have N in the range of $1024$ to $2048$ pixels. A lambertian BRDF with a single light source is used since we only want to show the performance of soft shadow calculation and not accurate surface shading.
\autoref{subfig:timen} shows that mean frame times are $5.4$~ms ($185$~Hz) for $N'=\{5,5,5\}$ (the default for most cases in this work) and $6.2$~ms  ($161$~Hz) for $N'=\{5,5,5,3\}$ which is only a $14.8$\% time penalty for a $20$\% increase in step count. By contrast, frame time is $18.2$~ms for uniform stepping (fixed 100 step count, step size $\Delta t=0.0006$), indicating that the speedup of our method is up to $237$\%. Another test on a slower laptop with Intel Core i5 CPU, 8 GB RAM, and integrated Intel Iris 540 GPU driving an internal 1920$\times$1080 display resulted in more modest, but still smooth frame rates of ${>}50$~Hz. See \autoref{tab:perfsummary} for a comparison of frame rates on each hardware at different display sizes (e.g., $512^2$ for machine learning).
\begin{table}[b]
    \caption{SELENE scene frame rates (Hz) for different hardware and window sizes. 1920$\times$1080 pixels is useful for interactive scene exploration while $512^2$ is more suitable for training machine learning systems.}%
    \label{tab:perfsummary}
    \centering
    \begin{tabular}{lcc}
        \hline
        GPU / Dimensions & 1920$\times$1080 & 512$\times$512 \\
        \hline
        GTX 970M & 185 & 402 \\
        Iris 540 & 50.5 & 196 \\
        \hline
    \end{tabular}
\end{table}


\autoref{subfig:timebreakdown} shows the proportion of total frame time taken up by shadow ray tracing.

Only visible height texture tiles are paged in synchronously on demand using a virtual texture scheme and stitched into a single texture to prevent visible seams between tiles.
Currently stitching is done by drawing tiles into a framebuffer object and frame time spikes coincide with stitching. More efficient methods such as asynchronous tile loading and blitting might reduce this.

%


Additional results are shown in \autoref{fig:a17} and \autoref{fig:other}. For Mars and Pluto, we reuse the existing Mie and Rayleigh atmospheric scattering model of celestia.Sci to modulate the final output of our raycasting fragment shader.

\section{Conclusion}
\label{sec:conclusion}

We presented a simple, novel method of efficiently rendering ray cast soft shadows on curved terrain by using maximum mipmaps computed in real-time and dynamic programming to find a global minimum shadow cost in constant runtime complexity. Additionally, we demonstrated a method of reducing view ray computation times using vertex pre-displacement to bootstrap ray starting positions. Combining these two methods, our ray casting engine runs in real-time with more than 200\% speed up over uniform ray stepping with comparable image quality and without hardware ray tracing acceleration. The ray casting engine is integrated into celestia.Sci, a general, interactive space simulation software. We demonstrated the scalability of our engine by generating lunar imagery across a wide range of distance scales, and accurately reproduced real lunar mission imagery containing heavily shadowed scenes with long penumbra. Artifacts due to approximating maximum delta heights along ray trajectories with a maximum mipmap were largely mitigated by starting at finer mip levels and concatenating several runs of the algorithm to cover long shadows.
Image accuracy and system performance remain quite acceptable given that we are running on previous-generation, commodity computing hardware.

    We thank Andrew Tribick for bug reports and feedback. This work was supported in part by funding from the Korea Ministry of Science and ICT through the ``Development of Pathfinder Lunar Orbiter and Key Technologies for the Second Stage Lunar Exploration'' project.

\appendix
\appendixpage

\theoremstyle{definition}
\newtheorem*{definition}{Definition}

\begin{definition}[Deriving $T$]
\label{def:deltaT}

Letting $\mathbf{R}$ be the object space position vector of the shadow ray starting point%
, $\mathbf{R}_\text{end}$ the position vector corresponding to $\mathbf{R}$ projected (spherical distortion) onto the texture plane at $z=1$ and offset by $\Delta\mathbf{M}$ along the shadow ray, we have:
\begin{equation}
\mathbf{R} = \mathbf{p}+t_1\mathbf{\hat{L}}
\end{equation}
\begin{equation}
\mathbf{R}_\text{end} = \frac{\mathbf{R}}{\mathbf{R}^{\mathrlap{z}}} + 2\frac{<\mathbf{\hat{L}}^{xy},0>}{\|<\mathbf{\hat{L}}^{xy},0>\|}\Delta\mathbf{M}
\end{equation}
(The factor of $2$ is due to object space conversion) $T$ can be found by deriving the intersection point between the shadow ray $\mathbf{R}+\mathbf{\hat{L}}T$ and $\mathbf{R}_\text{end}$\cite{HillJr.1994} (\autoref{fig:maxstep}):
{    \setlength{\belowdisplayskip}{0pt}%
	\setlength{\belowdisplayshortskip}{0pt}%
	\begin{align*}
	\mathbf{a} &= \mathbf{\hat{L}} & \mathbf{b} &= \mathbf{R}_\text{end}  & \mathbf{c} &= -\mathbf{R}
	\end{align*}%
}%
{    \setlength{\abovedisplayskip}{0pt}
	\setlength{\abovedisplayshortskip}{0pt}%
	\begin{equation}
	T = \frac{(\mathbf{c}\times\mathbf{b})\cdot(\mathbf{a}\times\mathbf{b})}{\|\mathbf{a}\times\mathbf{b}\|^{\mathrlap{2}}}\label{eq:maxstep}
	\end{equation}%
}
\begin{figure}[htbp]
	\centering
	\includegraphics[width=0.27\textwidth]{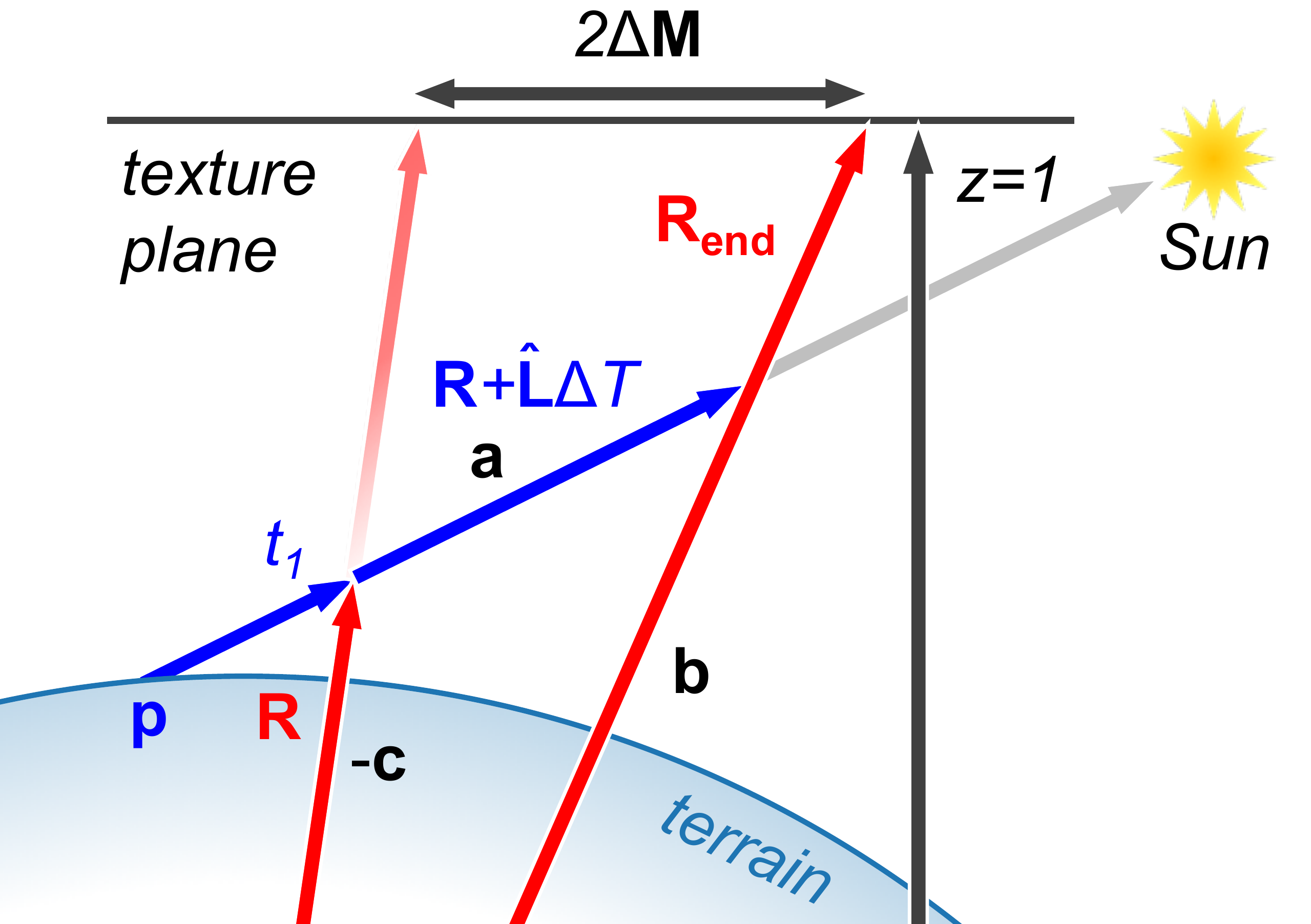}%
	\caption{Deriving $T$. The two lines (position vectors) $\mathbf{b}$ and $\mathbf{-c}$ are known to intersect at the origin, so we can use \autoref{eq:maxstep} for intersecting two lines in 3D to find the length $\mathbf{a}=T$.}\label{fig:maxstep}
\end{figure}
\end{definition}

\begin{figure*}[tp]
	\centering
	\begin{subfigure}{0.3\textwidth}
		\includegraphics[width=\textwidth]{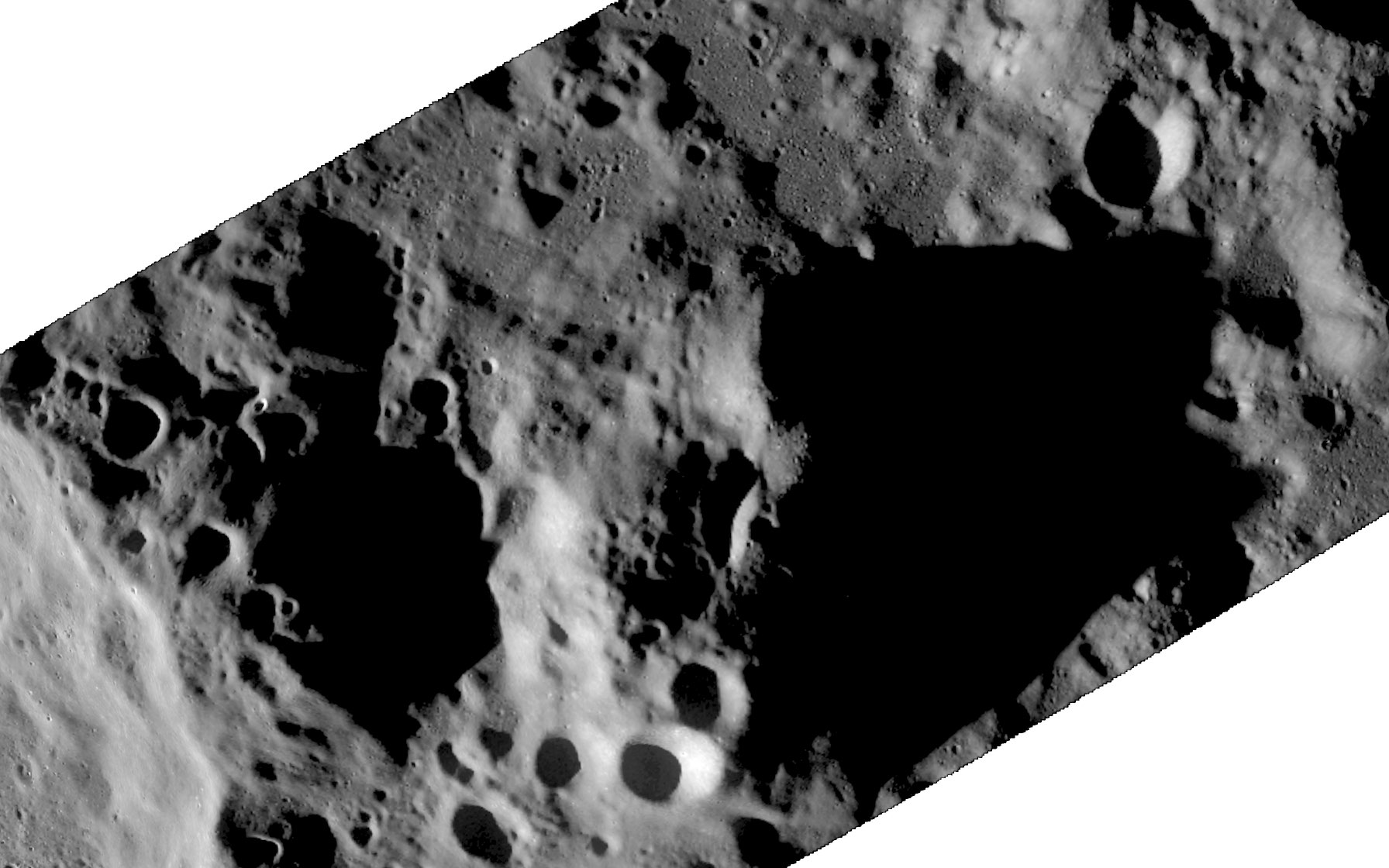}
	\end{subfigure}
	\begin{subfigure}{0.3\textwidth}
		\includegraphics[width=\textwidth]{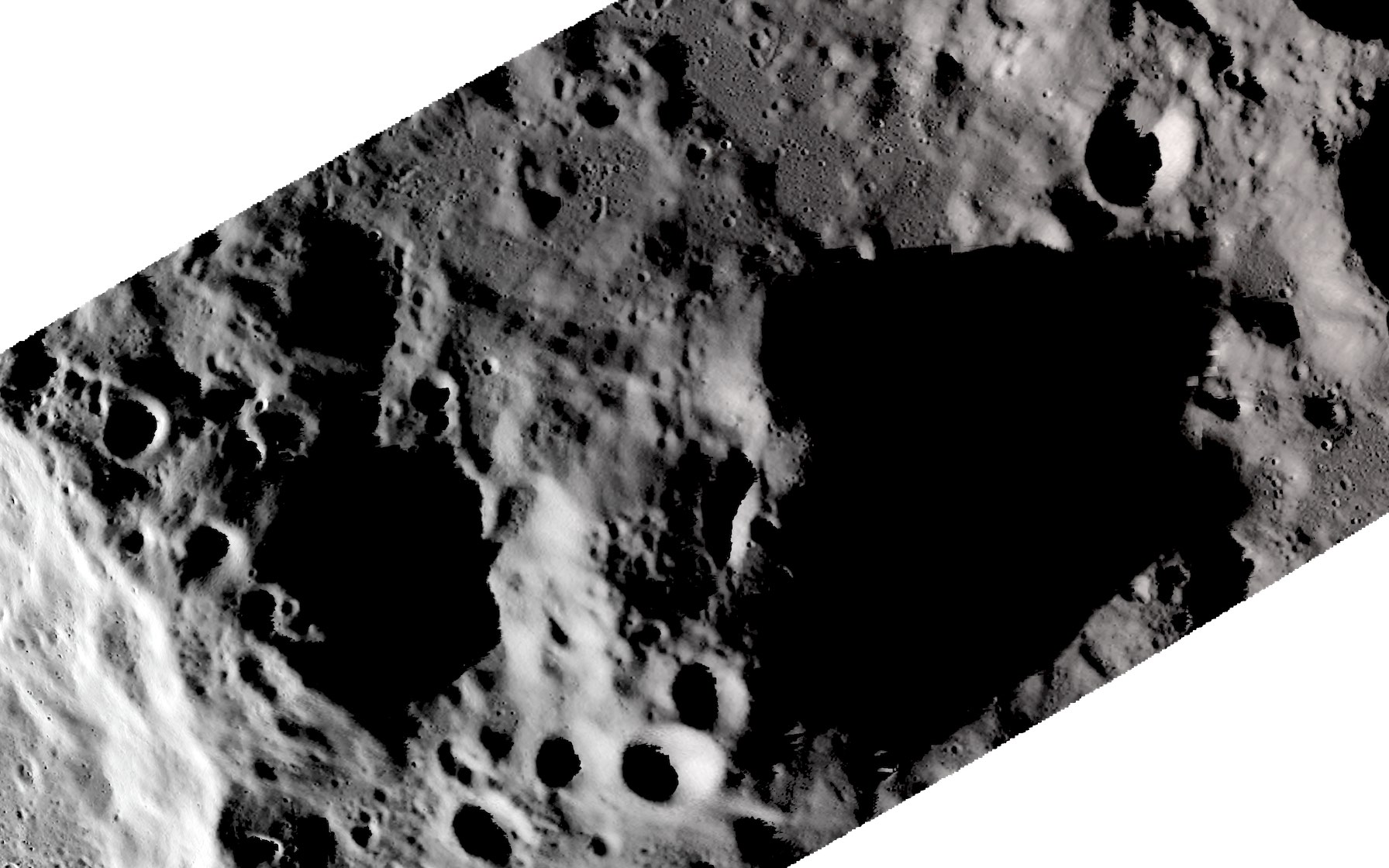}
	\end{subfigure}
	\begin{subfigure}{0.34\textwidth}
		\includegraphics[width=\textwidth]{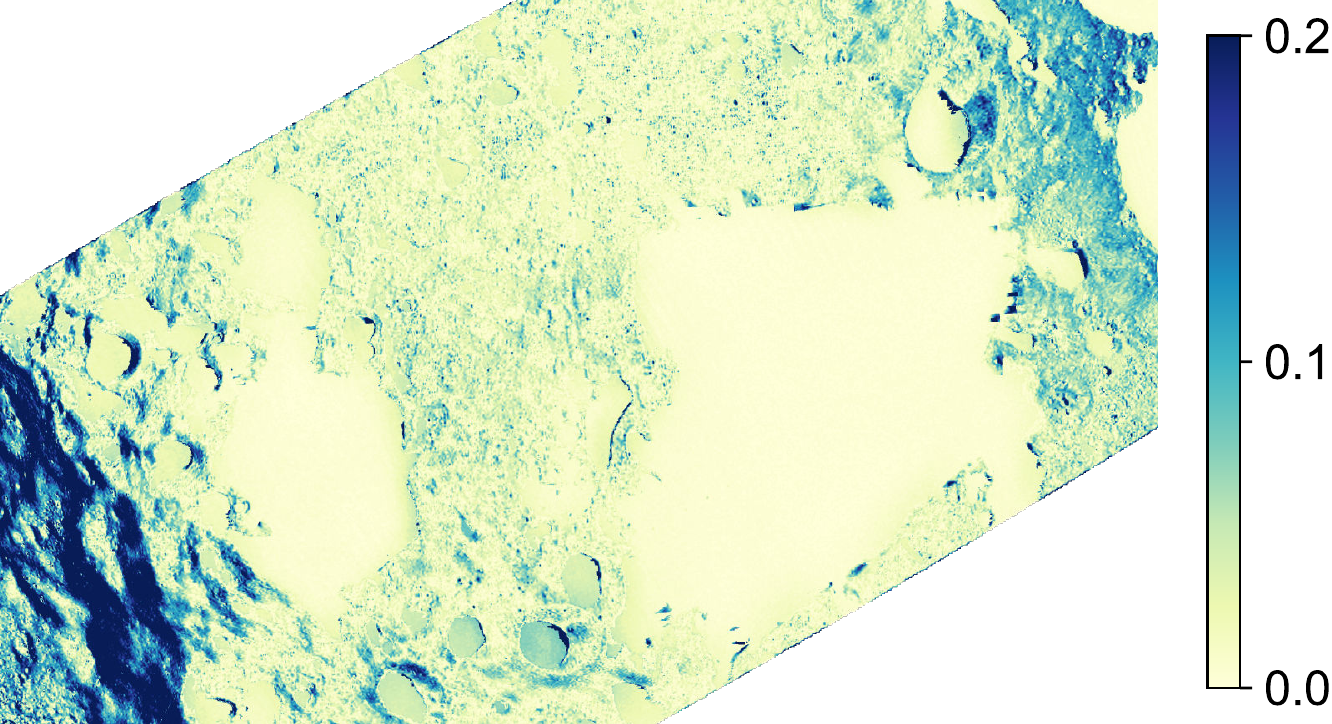}
	\end{subfigure}
	\\
	\vspace{1em}
	\begin{subfigure}{0.3\textwidth}
		\includegraphics[width=\textwidth]{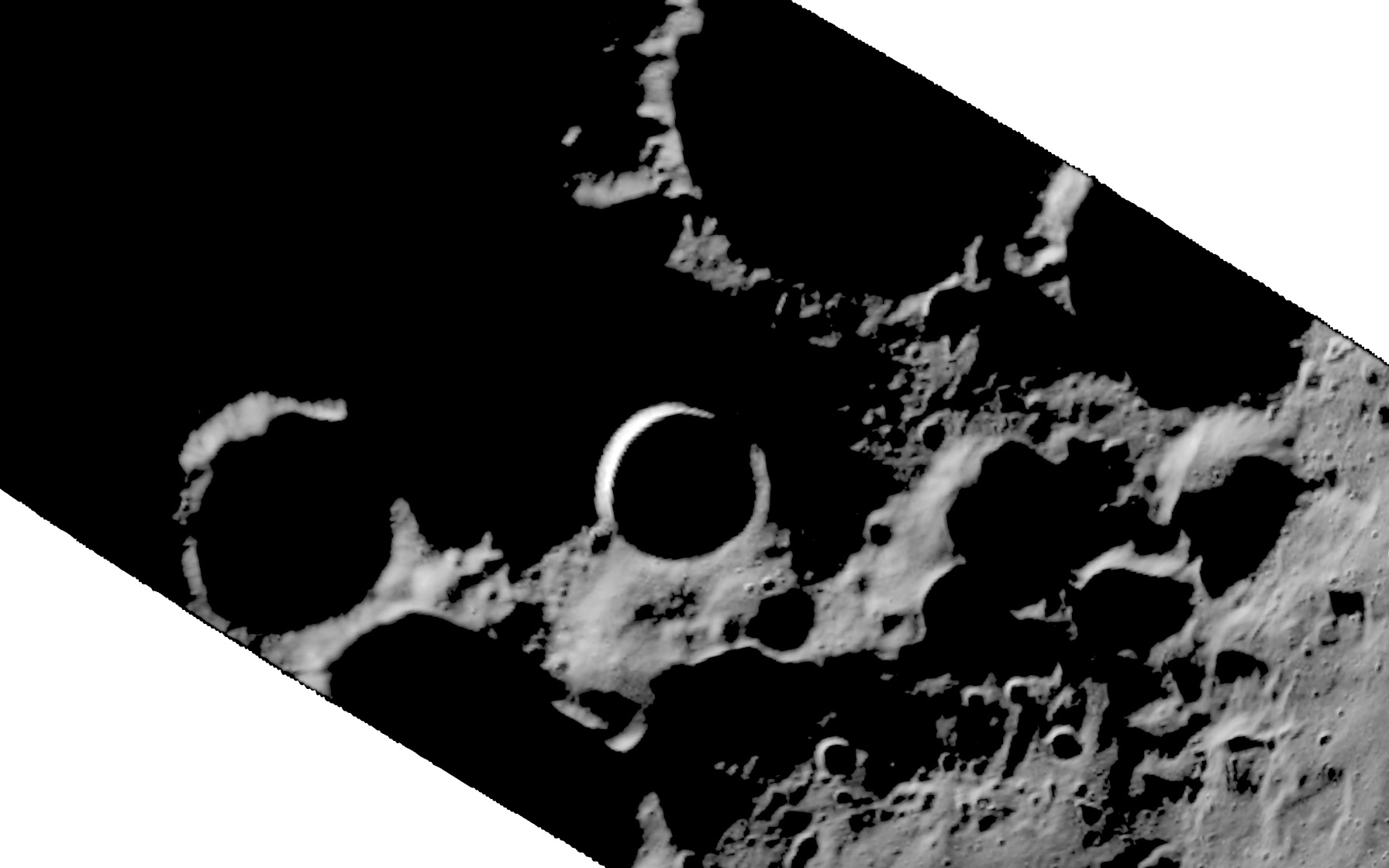}
	\end{subfigure}
	\begin{subfigure}{0.3\textwidth}
		\includegraphics[width=\textwidth]{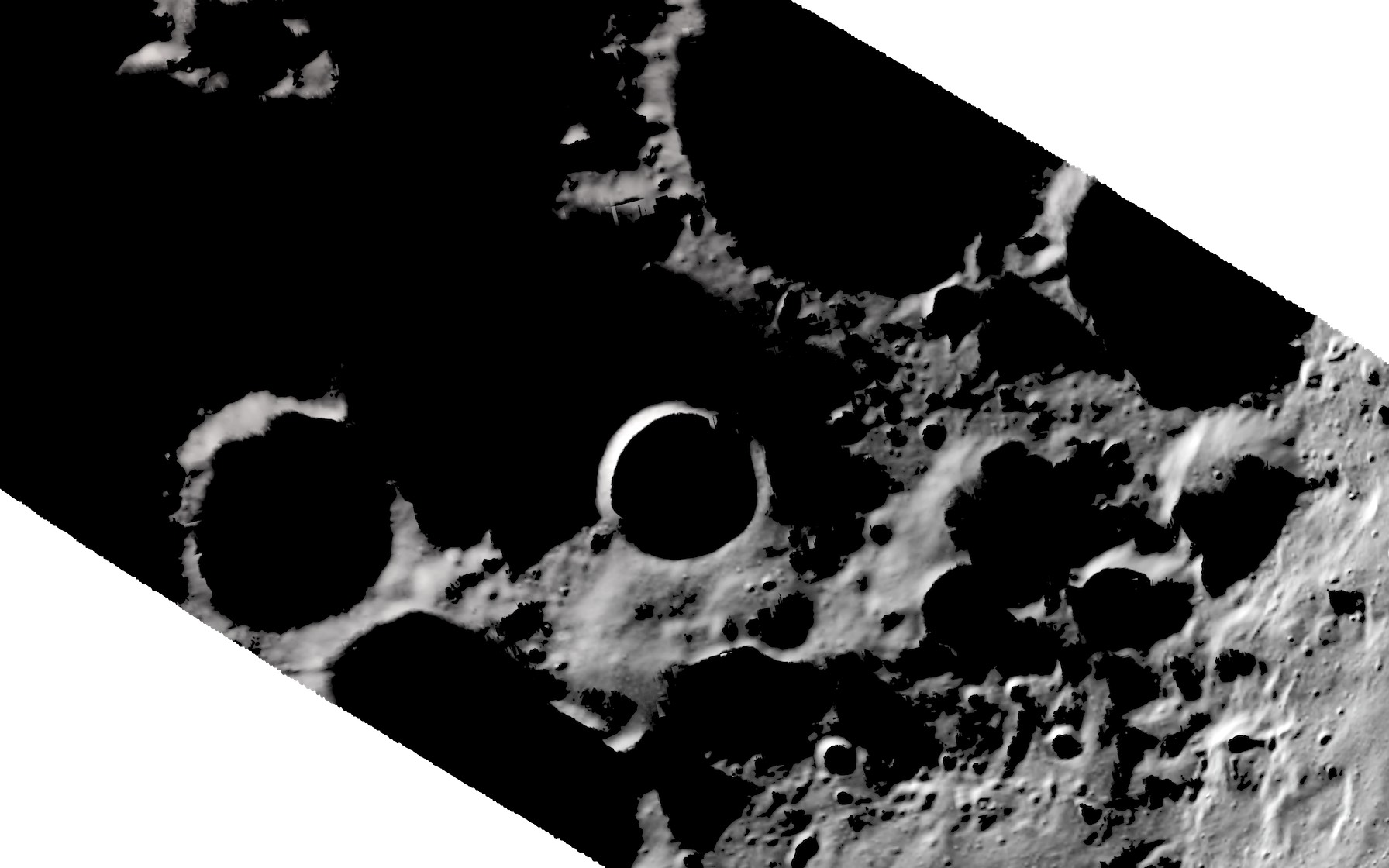}
	\end{subfigure}
	\begin{subfigure}{0.34\textwidth}
		\includegraphics[width=\textwidth]{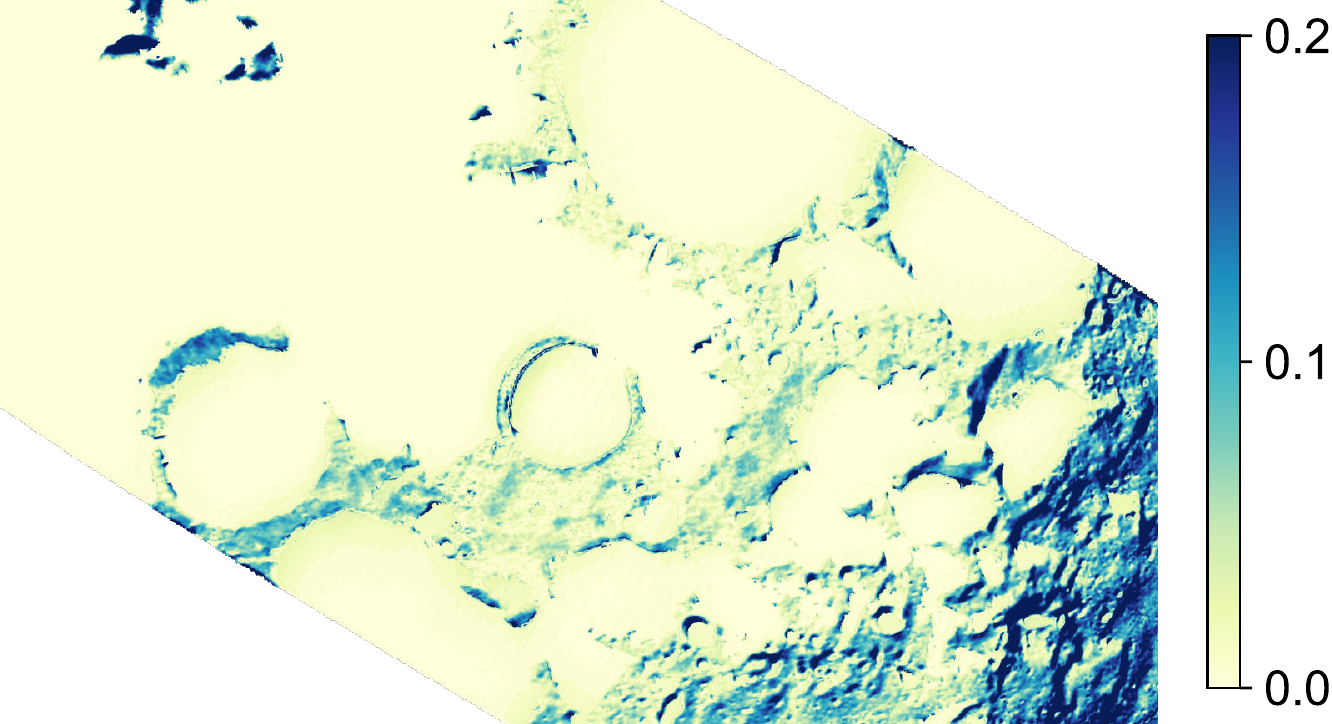}
	\end{subfigure}%
	\caption{Additional LRO WAC comparisons. Left column: reference images, center: our renders, right: absolute error. $N'={5,5,5,3}$ was used due to long shadow lengths.\label{fig:morepolarmovie}}
\end{figure*}
\begin{figure*}[p]
	\centering
	\hskip 2.9em%
	\begin{overpic}[trim=0cm 10cm 0cm 9cm,clip=true,width=0.3\textwidth]{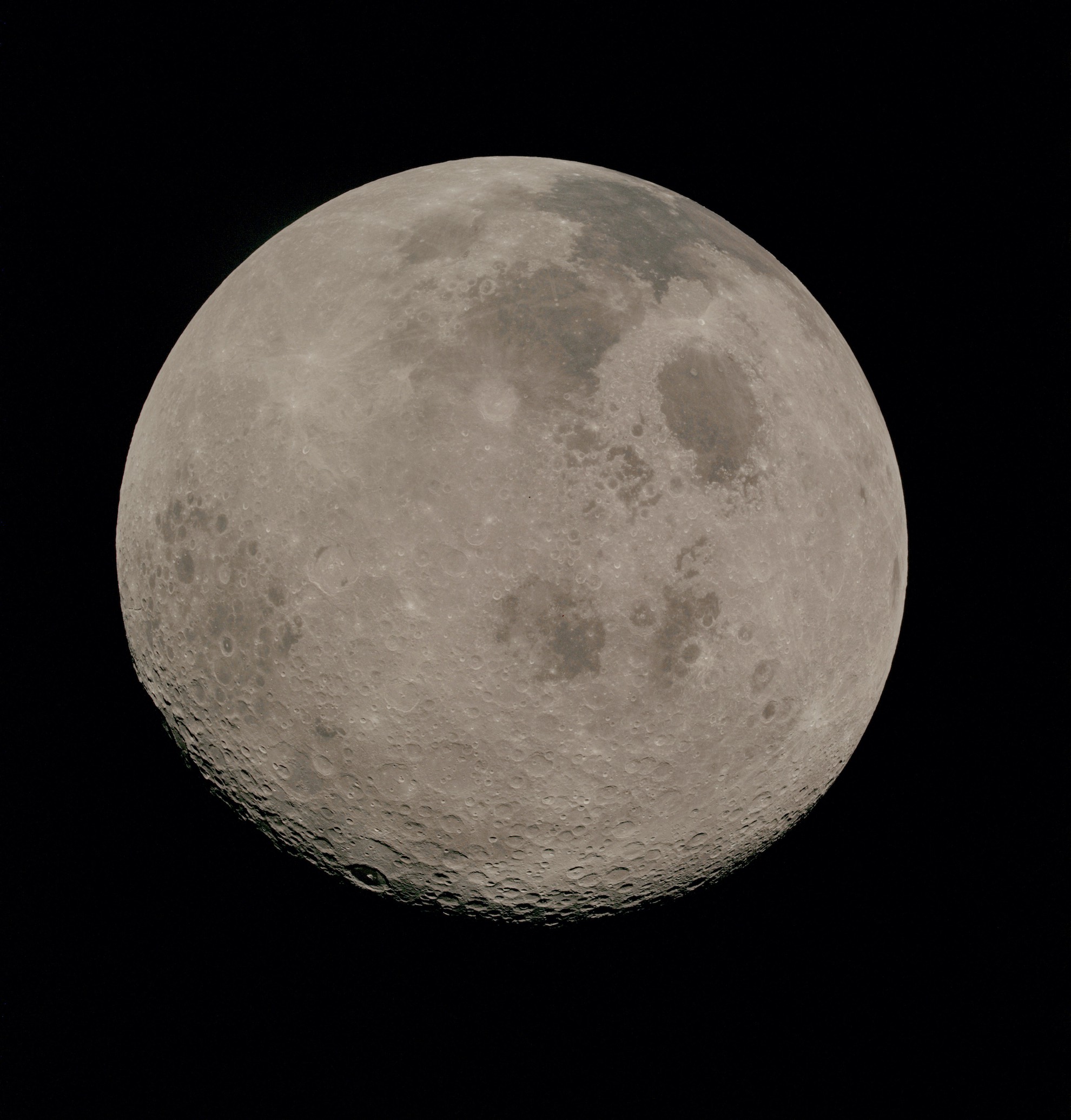}
		\put(3,67){\color{white}{(a)}}\phantomsubcaption{\label{subfig:a17photo}}
	\end{overpic}
	\begin{overpic}[trim=11.2cm 0cm 12.2cm 1.7cm,clip=true,width=0.3\textwidth]{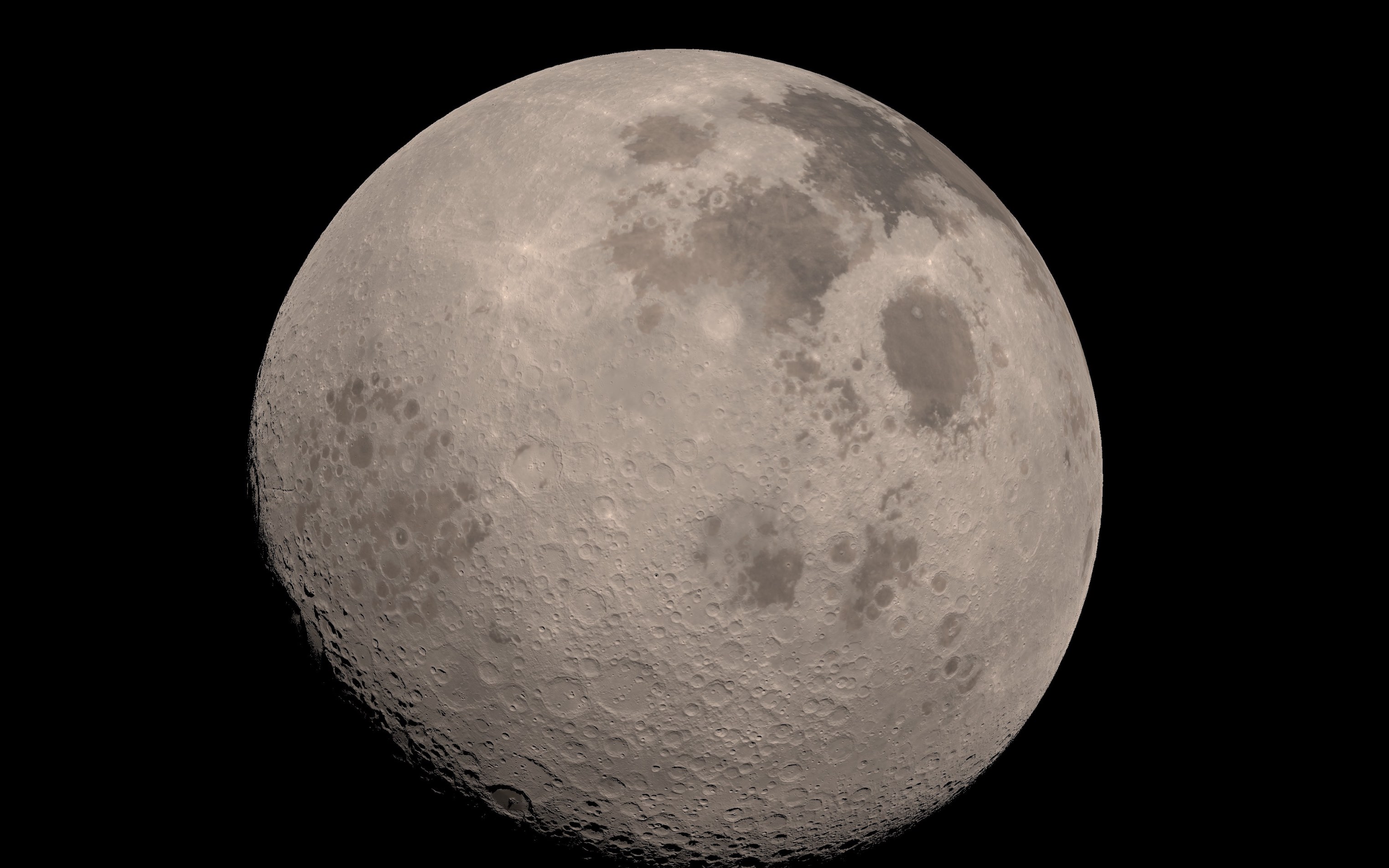}
		\put(3,67){\color{white}{(b)}}\phantomsubcaption{\label{subfig:a17render}}
	\end{overpic}
	\begin{overpic}[trim=25.0cm 0cm 52.0cm 54cm,clip=true,width=0.2485\textwidth]{AS17-render-wac_ilmenite4}
		\put(3,3){\color{white}{(b)}}\phantomsubcaption{\label{subfig:a17renderzoom}}
		\put(0,46){\includegraphics[trim=13cm 12cm 36cm 51cm,clip=true,width=0.2485\textwidth]{AS17-152-23311-hue_chroma2}}
		\put(3,53){\color{white}{(a)}}\phantomsubcaption{\label{subfig:a17photozoom}}
	\end{overpic}
	\caption{Apollo 17 photograph AS17-152-23311, color-matched (\subref*{subfig:a17photo}) vs our render (\subref*{subfig:a17render}).\label{fig:a17}}
\end{figure*}
\begin{figure*}[p]
	\centering
	\captionsetup{margin=1cm}
	\begin{overpic}[trim=2cm 7cm 2cm 7.5cm,clip=true,width=0.7\textwidth]{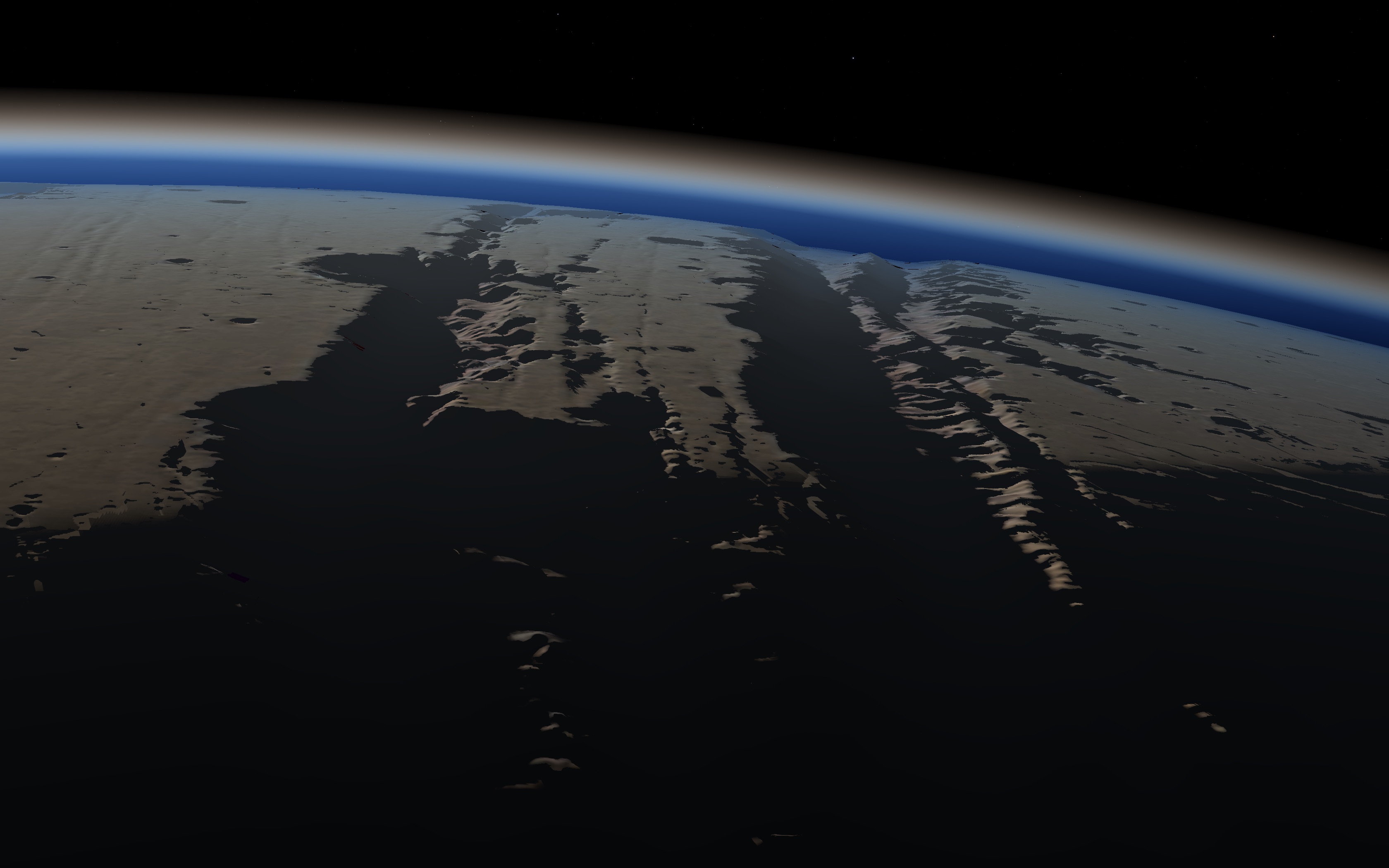}%
		\put(93,47){\color{white}{(a)}}
	\end{overpic}\\
	\begin{overpic}[trim=2cm 7cm 2cm 7.5cm,clip=true,width=0.7\textwidth]{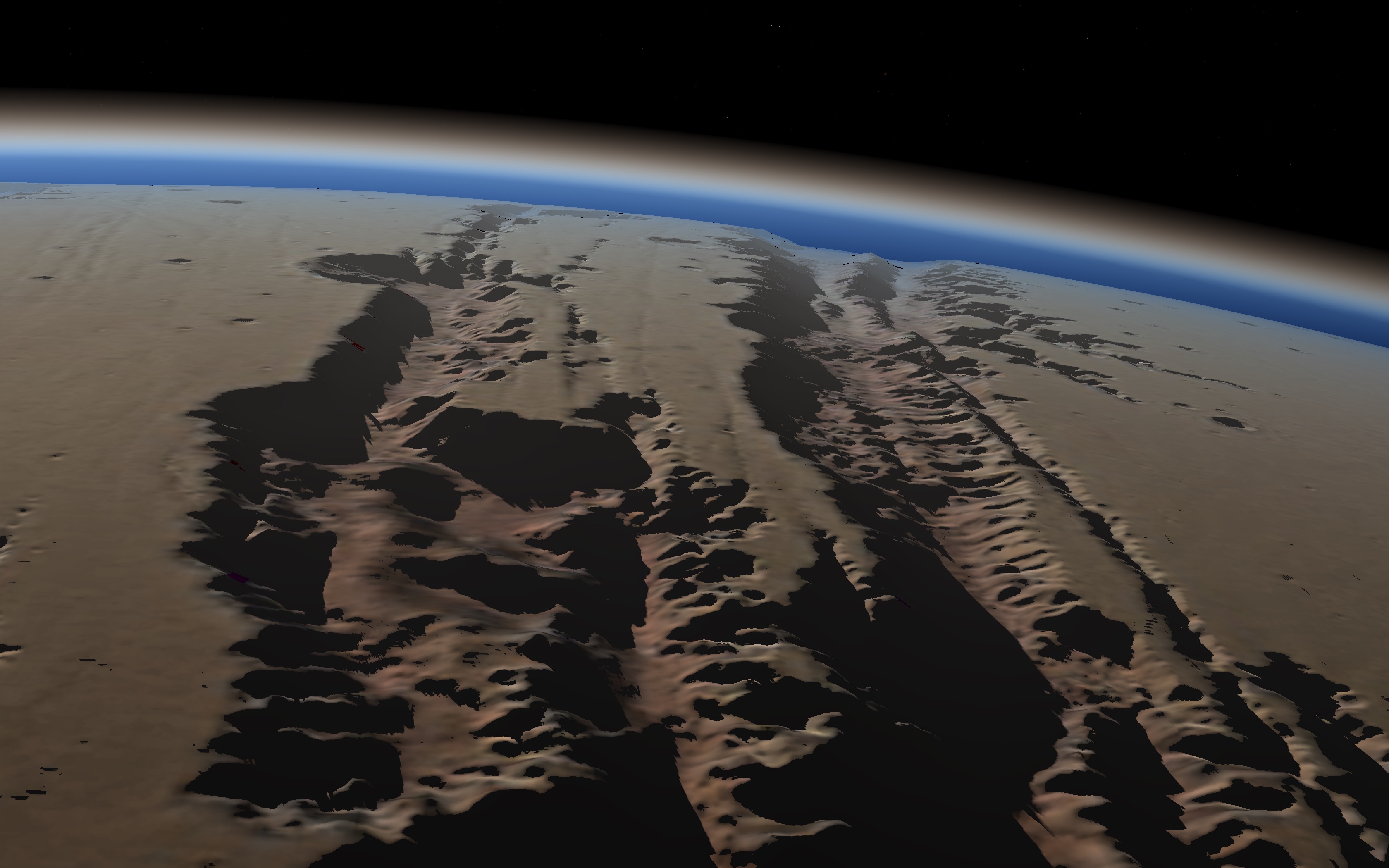}%
		\put(93,47){\color{white}{(b)}}
	\end{overpic}\\\vspace{2em}
	\begin{overpic}[trim=0cm 6cm 14cm 13.8cm,clip=true,width=0.7\textwidth]{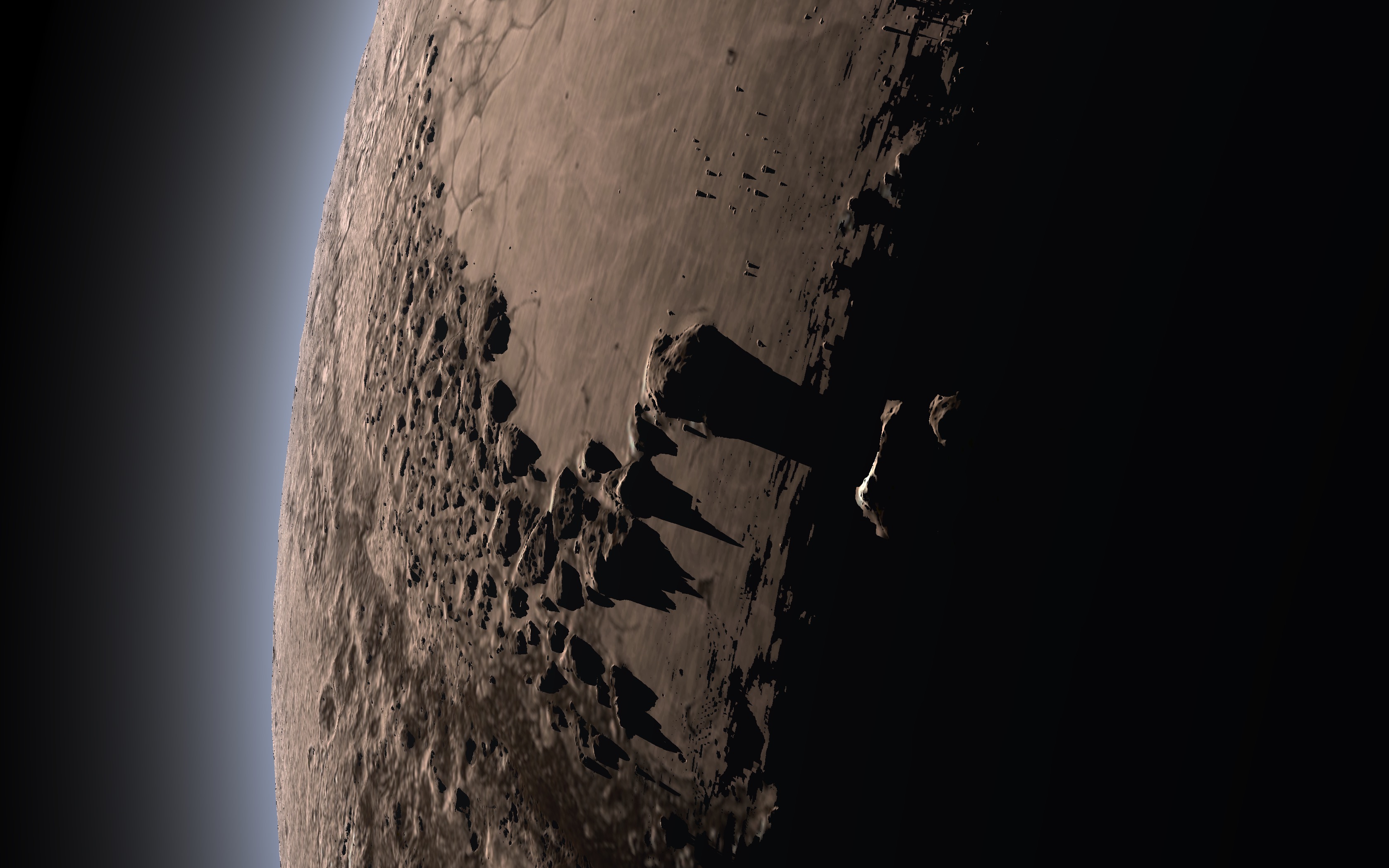}%
		\put(93,47){\color{white}{(c)}}
	\end{overpic}
	\caption{Valles Marineris on Mars (a), (b) 
		rendered using Mars Orbiter Laser Altimeter 463m/px height map\cite{Neumann2003}. Sputnik Planitia on Pluto (c) rendered with 300m/px height map from New Horizons data\cite{Schenk2018}.\label{fig:other}}
\end{figure*}


\clearpage
\bibliographystyle{plain}
\bibliography{moon}

\end{document}